\documentclass[lettersize,journal]{IEEEtran}

\usepackage{subcaption}
\usepackage{graphicx}
\usepackage{grffile}
\usepackage{lipsum}
\usepackage[symbol]{footmisc}
\usepackage{amsmath,amssymb,amsfonts}
\usepackage{textcomp}
\usepackage{xcolor}
\usepackage{cite}

\def\BibTeX{{\rm B\kern-.05em{\sc i\kern-.025em b}\kern-.08em
    T\kern-.1667em\lower.7ex\hbox{E}\kern-.125emX}}

\usepackage[linesnumbered,commentsnumbered,ruled,vlined]{algorithm2e}
\usepackage{caption}
\usepackage{booktabs}
\usepackage{xcolor, colortbl}
\usepackage{amsmath}
\usepackage{array}
\usepackage{booktabs}
\usepackage[utf8]{inputenc}
\usepackage{algorithmic}
\usepackage{adjustbox}
\usepackage{float}

\begin{document}
\newcommand{\best}[1]{\cellcolor{green!30} #1}
\newcommand{\worst}[1]{\cellcolor{red!30} #1}
\newcommand{\up}{\(\uparrow\)}
\newcommand{\down}{\(\downarrow\)}
\renewcommand{\thefootnote}{\fnsymbol{footnote}}
\newcommand{\tk}[1]{\textcolor{red} {[TK: #1]}}
\newcommand{\zulkar}[1]{\textcolor{purple}{[ZN: #1]}}

\title{Energy-Efficient and High-Performance \\Data Transfers with DRL Agents}


\author{
    Hasibul Jamil,~\IEEEmembership{Student Member,~IEEE}, Jacob Goldverg, Elvis Rodrigues, \\MD S Q Zulkar Nine,~\IEEEmembership{Member,~IEEE}, and Tevfik Kosar,~\IEEEmembership{Senior Member,~IEEE}
    \IEEEcompsocitemizethanks{
        \IEEEcompsocthanksitem Hasibul Jamil, Jacob Goldverg, Elvis Rodrigues, and Tevfik Kosar are with the Department of Computer Science and Engineering, University at Buffalo, Buffalo, NY, 14260.\protect\\
        E-mail: \{mdhasibu, jacobgol, elvisdav, tkosar\}@buffalo.edu
        \IEEEcompsocthanksitem MD S Q Zulkar Nine is with the Department of Computer Science, Tennessee Technological University, Cookeville, TN, 38505.\protect\\
        E-mail: mnine@tntech.edu
    }
}

\IEEEtitleabstractindextext{%
\begin{abstract}

The rapid growth of data across fields of science and industry has increased the need to improve the performance of end-to-end data transfers while using the resources more efficiently. In this paper, we present a dynamic, multiparameter deep reinforcement learning (DRL) framework that adjusts application-layer transfer settings during data transfers on shared networks. Our method strikes a balance between high throughput and low energy utilization by employing reward signals that focus on both energy efficiency and fairness. The DRL agents can pause and resume transfer threads as needed, pausing during heavy network use and resuming when resources are available, to prevent overload and save energy. We evaluate several DRL techniques and compare our solution with state-of-the-art methods by measuring computational overhead, adaptability, throughput, and energy consumption. Our experiments show up to 25\% increase in throughput and up to 40\% reduction in energy usage at the end systems compared to baseline methods, highlighting a fair and energy-efficient way to optimize data transfers in shared network environments.


\end{abstract}

}

\maketitle
\thispagestyle{empty}

\IEEEdisplaynontitleabstractindextext
\IEEEpeerreviewmaketitle

\section{Introduction}

\IEEEPARstart{T}{he} explosive growth of data generated by scientific research, industrial applications, e-commerce, social networks, the Internet of Things (IoT), and large-scale AI training workloads has driven global data traffic past 5 zettabytes per year as of 2023~\cite{Cisco2023}—equivalent to shipping over 3 billion DVDs daily. Accessing, sharing, and disseminating data at this scale in an efficient and sustainable manner remains an open challenge.

A critical dimension of this challenge is energy consumption. Information and communication technologies are projected to consume between 8\% and 21\% of global electricity by 2030~\cite{belkhir2018assessing}, with communication networks alone accounting for roughly 43\% of total IT power draw~\cite{andrae2015global}. Internet data transfers already consume over a hundred terawatt-hours annually, costing approximately 20 billion US dollars~\cite{nine2018greendataflow}, and network energy costs already represent 40--50\% of operational expenditures for telecommunications operators in developing countries~\cite{lorincz2019greener}. Recent analyses warn that the cost of data movement already dominates electrical losses and may undermine future gains in compute energy efficiency if left unaddressed~\cite{shalf2020future}. In some scenarios, physically shipping storage media is less carbon-intensive than network transfer~\cite{aujoux2021estimating}.

Despite extensive work on power management for networking infrastructure, energy consumption at the \emph{end systems} (sender and receiver nodes) during active transfers has received little attention. Yet end-system power can account for approximately 25\% of total transfer energy on intercontinental paths, rising to 60\% on nationwide networks and up to 90\% on local area networks~\cite{powerModelAlan}. Because this ratio depends on the number of intermediate network devices and their per-device power draw, reducing end-system consumption can yield substantial savings.

Achieving these savings requires parameter adaptation strategies that cope with highly dynamic conditions—fluctuating bandwidth, varying round-trip times, and competing cross-traffic. Traditional heuristic approaches lack the robustness to generalize across such diverse scenarios. To address this, we propose \texttt{SPARTA} (\underline{S}mart \underline{P}arameter \underline{A}daptation via \underline{R}einforcement Learning for data \underline{T}ransfer \underline{A}cceleration), a deep reinforcement learning (DRL) framework with the following contributions:

\begin{itemize}
    \item  We train DRL agents with reward signals that balance energy consumption against fairness, enabling agents to dynamically pause and resume transfer threads based on real-time conditions to prevent resource oversaturation and reduce energy usage.

    \item  By jointly adjusting transfer parameters, agents avoid overloading bandwidth and CPU cores during peak periods while exploiting available resources during off-peak periods, accelerating transfers while preserving fairness among concurrent users.

    \item  We introduce an emulation environment that replays state-transition logs from early real-world episodes, enabling efficient policy learning without the prohibitive time and energy costs of prolonged live transfers.

    \item  Experiments demonstrate up to 25\% throughput improvement and up to 40\% reduction in energy consumption compared to baseline methods.
\end{itemize}

The remainder of the paper is organized as follows: Section~II provides background and related work on energy-efficient high-performance data transfers. Section~III details the design and implementation of the proposed framework. Section~IV presents experimental evaluations, and Section~V concludes with insights and future directions.

\vspace{-2mm}
\section{Background and Related Work}

\begin{figure*}[ht]
    \centering
    \includegraphics[width=\linewidth]{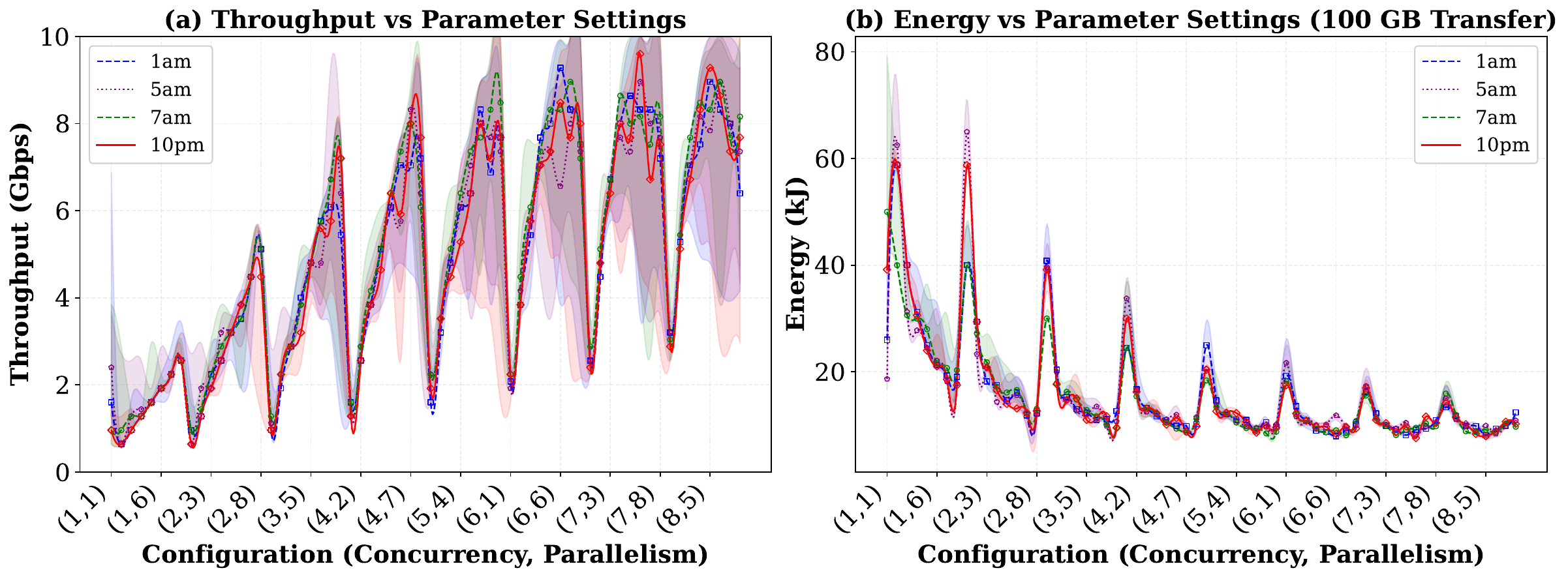}
    \caption{\small The plots show the throughput and energy consumption for different concurrency and parallelism settings under varying background traffic conditions observed at different times of the day in the Chameleon Cloud between TACC and UC sites, connected by a 10\,Gbps network link. Subfigure~(a) highlights the throughput behavior, while subfigure~(b) shows the  energy consumption both influenced by the chosen levels of concurrency and parallelism. These results demonstrate the performance and energy efficiency trade-offs under different background traffic}
    \label{figure:th-en-parameter}
\end{figure*}

\subsection{Application-Layer Parameter Tuning}

Improving data transfer performance at the transport layer—e.g., via new congestion control protocols~\cite{DataCenterCongestionControl-2022, qtcp, BBR2016, PCC-Vivace}—requires kernel and system modifications that hinder wide adoption. A more practical strategy is tuning application-layer parameters such as pipelining ($pp$), parallelism ($p$), concurrency ($cc$), buffer size, and striping~\cite{Yildirim-TCC2016}, of which parallelism and concurrency are the most impactful~\cite{balman2007data, Yildirim-TCC2016, kim2015highly}.

{\em Concurrency ($cc$)} is task-level parallelism: multiple server processes or threads transfer different files simultaneously. {\em Parallelism ($p$)} is the number of TCP streams per file transfer, primarily beneficial for medium and large files. Increasing the total stream count ($cc \times p$) improves throughput, but excessive streams provoke TCP congestion avoidance and reduce the sending rate. When properly tuned, these parameters address common bottlenecks such as inadequate I/O parallelism and TCP buffer limitations~\cite{jamil-2023-tcp}.
\vspace{-3mm}
\subsection{Throughput Optimization}

Prior approaches to application-layer tuning fall into three categories. {\em Heuristic-based} systems~\cite{arslan2018big, balman2007data, di2019cross, blaze, skyplane, effingo, kosar2005data, arslan2018high} outperform single-stream baselines but generalize poorly to dynamic network conditions. {\em Supervised learning} methods~\cite{jamil-ICCCN-2022, Rodolph2021, twoPhase} predict optimal settings from historical data, yet assembling large, representative datasets under diverse conditions is expensive. {\em Online} methods~\cite{Dong-2005, Yildirim-TCC2016, falcon-2023} adapt continuously but suffer from slow convergence and lack fairness safeguards in multi-tenant environments.

Widely used tools reflect these limitations. Globus~\cite{globus} applies static concurrency (2--8) and cannot react to changing conditions, often underutilizing available bandwidth. ProbData~\cite{probData} uses stochastic approximation but can require hours to converge, while PCP~\cite{Yildirim-TCC2016} employs hill-climbing over a reduced search space, limiting both speed and precision. Other work~\cite{balman2007data, twoPhase} relies on heuristic or historical-data models whose dependence on extensive production-network logs makes them impractical in many settings.
\vspace{-3mm}
\subsection{Energy Optimization}

Most energy-efficiency research targets the network core—switches and routers~\cite{Brooks:2000:WFA:339647.339657, rawson2004mempower, zedlewski2003modeling, gurumurthi2002using, contreras2005power, economou2006full, fan2007power, rivoire2008comparison, koller2010wattapp, hasebe2010power, vrbsky2013decreasing, Katz_2008, Mahadevan_2009, Greenberg_2009, nine2023greennfv}. These techniques reduce network-level consumption but can degrade performance (e.g., by sleeping idle components) or require costly hardware upgrades, limiting near-term practicality.

Far fewer studies address energy consumption at the end systems during active transfers. Existing approaches—heuristic-based~\cite{alan2015energy, di2019cross, guner2018energy} and historical-data-driven~\cite{Rodolph2021, jamil-ICCCN-2022}—adjust transfer parameters to meet throughput or energy targets specified by SLAs. However, building robust historical models demands extensive data across diverse traffic conditions, and heuristics generalize poorly to unseen network settings.


In our prior work~\cite{jamil-2023-tcp}, we tune TCP stream counts for throughput optimization. \texttt{SPARTA} extends this by adopting a multi-parameter DRL framework that jointly optimizes concurrency ($cc$) and parallelism ($p$) while incorporating energy efficiency and fairness directly into the reward function, ensuring transfers that are faster, more energy-efficient, and fair in shared network environments.

\newlength\myindent
\setlength\myindent{5em}
\newcommand\bindent{%
    \begingroup
    \setlength{\itemindent}{\myindent}
    \addtolength{\algorithmicindent}{\myindent}
}
\newcommand\eindent{\endgroup}
\vspace{-2mm}
\section{SPARTA Model Overview}
\label{sec:sparta_overview}

In \texttt{SPARTA}, we use Reinforcement Learning (RL) effectively for balancing the energy consumption during data transfers with the achieved throughput and fairness. 
In the following subsections, we discuss why we choose RL and the details of our RL solution.  
\vspace{-3mm}
\subsection{Why Reinforcement Learning?}
Existing heuristic solutions cannot easily adapt to the dynamic nature of wide-area networks, where optimal concurrency ($cc$) and parallelism ($p$) depend on both static factors (e.g., hardware limits, available network bandwidth) and highly variable conditions (e.g., background traffic, transient congestion). 
Figure~\ref{figure:th-en-parameter} displays throughput and energy consumption for various concurrency and parallelism settings under different network traffic conditions in the Chameleon Cloud~\cite{keahey2020lessons} between TACC and UC sites. These sites are connected by a 10 Gbps network, and the experiment involved transferring 100 files of 1 GB each using TCP CUBIC~\cite{cubic2008}. The plots illustrate how varying the levels of $cc$ and $p$ influences receiver-side throughput and energy usage. The results highlight the trade-offs between performance and energy efficiency under various network conditions. The relationship between the parameters is non-linear, and the optimal settings can improve performance by up to 10 times compared to baseline settings (where concurrency ($cc$) = 1 and parallelism ($p$) = 1). As background traffic changes, the optimal settings for throughput and energy also shift. With limited network signals available from end hosts, an ideal data transfer solution needs to adjust $cc$ and $p$ to maintain optimal performance and energy consumption.

To motivate the need for adaptive control, we recall classic loss-based TCP throughput behavior (TCP CUBIC in our experiments). When packet loss ratio $L<1\%$, a single TCP flow throughput can be approximated~\cite{mathis1997} as:
\begin{equation}
TCP_{\mathrm{thr}} \lesssim \frac{MSS}{RTT}\cdot \frac{C}{\sqrt{L}},
\label{eq:single_tcp}
\end{equation}
and for $n$ parallel TCP streams, the aggregate throughput can be approximated by summing per-flow rates~\cite{hacker2002} as:
\begin{equation}
TCP_{\mathrm{agg}} \lesssim \frac{C}{RTT}\sum_{i=1}^{n}\frac{MSS}{\sqrt{L_i}}.
\label{eq:multi_tcp}
\end{equation}
Increasing streams can increase aggregate throughput until the bottleneck saturates. Beyond that point, congestion increases loss and round trip time (RTT), which reduces goodput and increases energy waste (retransmissions and waiting). Because the bottleneck state changes over time, heuristics with fixed or slowly adapting parameters are not ideal.

We therefore model this adaptive control as a sequential decision problem and use Deep Reinforcement Learning (DRL) to learn a policy that maps end-host signals to parameter updates, without requiring prior knowledge of network internals.

\subsection{Optimization Goals and Formulation}
\label{subsec:objectives}

At each monitoring interval (MI) $t$, the controller chooses concurrency and parallelism $(cc_t,p_t)$ and creates a TCP stream budget:
\begin{equation}
n_t \triangleq cc_t\cdot p_t,\qquad n_t\in\{1,2,\dots,N\},
\label{eq:stream_budget}
\end{equation}
where $N$ is a safe number of TCP streams denoting end-host and network stability constraints.

Let $T_t$, $L_t$, and $E_t$ be the measured throughput, packet loss rate, and end-system energy during MI $t$.
\texttt{SPARTA} aims to maximize data throughput while (1) promoting fairness and indirectly reducing energy via low congestion, or (2) prioritizing throughput per unit energy.

\begin{figure*}[t]
  \centering
  \begin{subfigure}[b]{0.40\textwidth}
    \includegraphics[width=\textwidth]{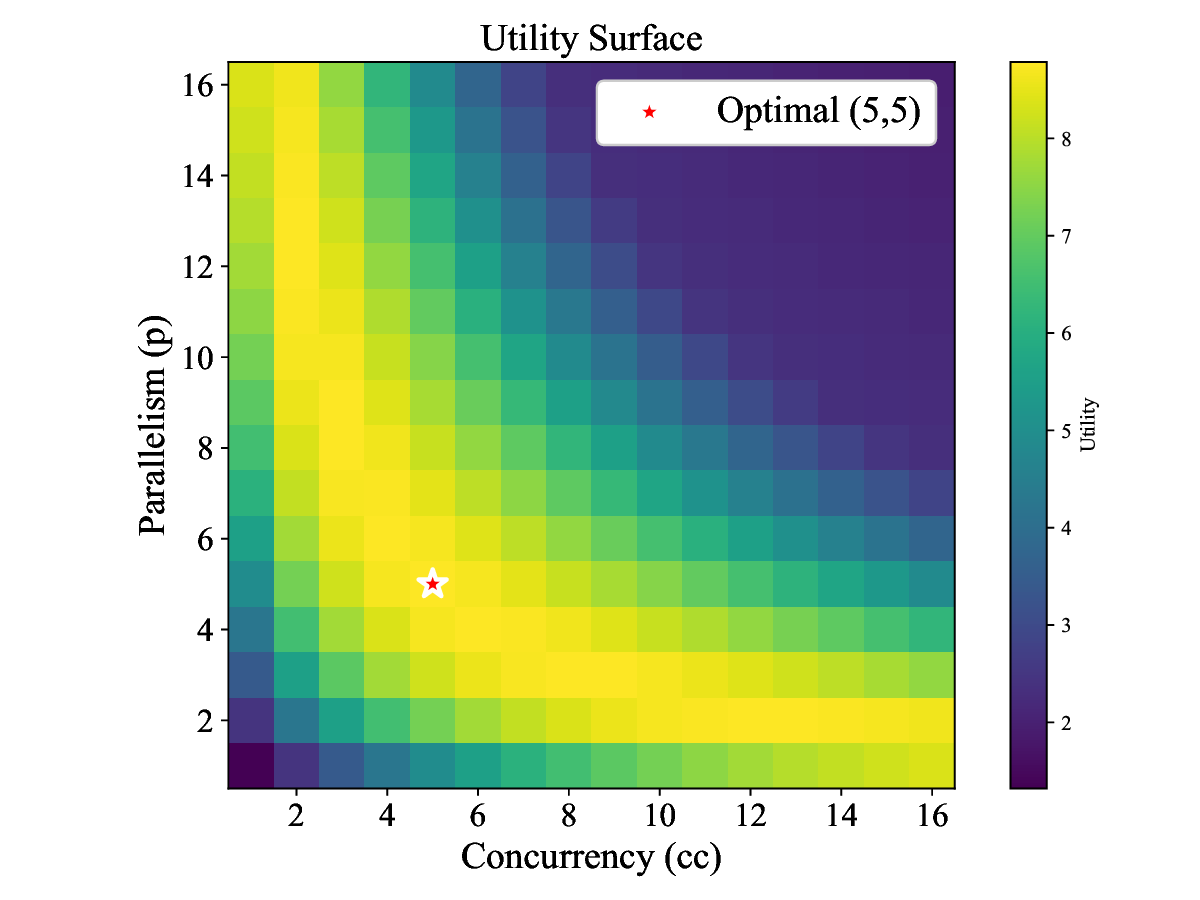}
    \caption{Utility (Fairness \& Efficiency) Surface over $(cc,p)$}
    \label{fig:utility_surface}
  \end{subfigure}
   \begin{subfigure}[b]{0.40\textwidth}
    \includegraphics[width=\textwidth]{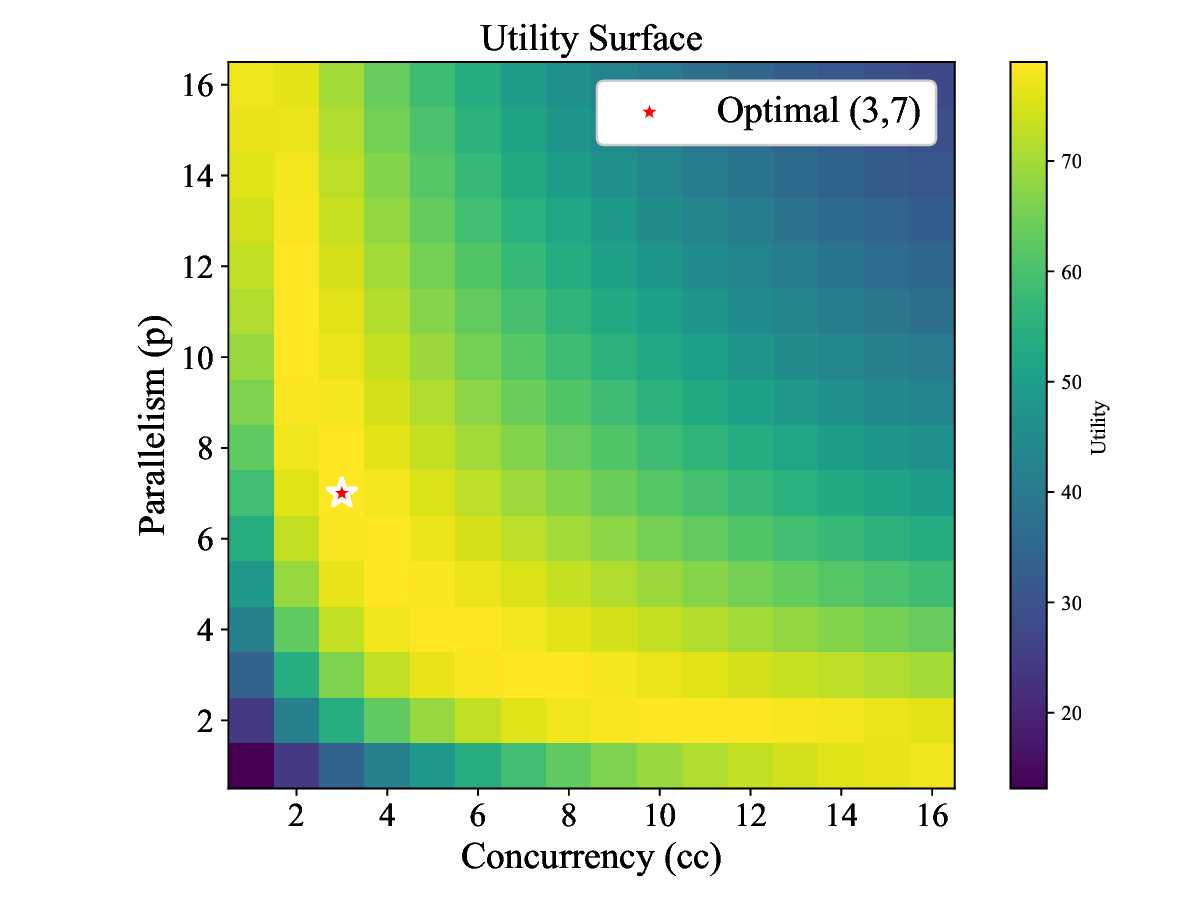}
    \caption{Utility (Throughput/Energy) Surface over $(cc,p)$ }
    \label{fig:utility_surface_te}
  \end{subfigure}
\caption{\small Utility geometry and trackability.
  (a) and (b) Utility induced by Eq.~\eqref{eq:utility_fe} and Eq.~\eqref{eq:utility_te} over the feasible $(cc,p)$ grid exhibits a single dominant peak near the knee region, consistent with a unimodal/quasi-concave profile in $n=cc\cdot p$ for a fixed network condition.}
  \label{fig:theoretical_validation}
\end{figure*}

\subsubsection{Fairness and Efficiency Objective (SPARTA-FE)}
In shared networks, high throughput should not be achieved by pushing the bottleneck into persistent congestion that harms other transfers.
We adopt a utility function ~\cite{falcon-2023} that rewards throughput while penalizing congestion (loss):
\begin{equation}
U_{\mathrm{FE}}(T_t,L_t,n_t) \triangleq \frac{T_t}{K^{n_t}} - B\,T_t\,L_t,
\label{eq:utility_fe}
\end{equation}
where $K>1$ and $B>0$ are constants controlling the relative weight of throughput scaling and the cost of loss. By discouraging congestion (i.e., excessive packet loss), this objective not only ensures fairness but also reduces retransmissions and idle waiting, indirectly improving energy efficiency.

Over a transfer session $[t_s,t_f]$, the objective is to maximize cumulative utility:
\begin{equation}
\max_{\{cc_t,p_t\}} \sum_{t=t_s}^{t_f} U_{\mathrm{FE}}(T_t,L_t,n_t)
\quad \text{s.t.}\quad 1\le n_t\le N,\; T_t\le b,
\label{eq:fe_objective}
\end{equation}
where $b$ is the upper bound on achievable throughput.

\subsubsection{Throughput-Focused Energy Objective (SPARTA-T)}
For energy-sensitive deployments, we optimize throughput per unit energy:
\begin{equation}
U_{\mathrm{TE}}(T_t,E_t) \triangleq \frac{T_t}{E_t},
\label{eq:utility_te}
\end{equation}
and maximize
\begin{equation}
\max_{\{cc_t,p_t\}} \sum_{t=t_s}^{t_f} U_{\mathrm{TE}}(T_t,E_t)
\quad \text{s.t.}\quad 1\le n_t\le N,\; T_t\le b.
\label{eq:te_objective}
\end{equation}

\subsubsection{Unifying View}
Both objectives are implemented through a selectable reward:
\begin{equation}
r_t =
\begin{cases}
U_{\mathrm{FE}}(T_t,L_t,n_t), & \text{(Fairness \& Efficiency)},\\[1mm]
U_{\mathrm{TE}}(T_t,E_t), & \text{(Throughput/Energy)}.
\end{cases}
\label{eq:reward_def}
\end{equation}
By penalizing loss, the first reward indirectly lowers energy overhead through congestion avoidance, while the second directly optimizes throughput per unit energy. In practice, either reward can be selected based on deployment needs, allowing \texttt{SPARTA} to accommodate diverse operational goals and constraints.
\vspace{-5mm}
\subsection{Theoretical Validation: Solvability and Utility Geometry}
\label{subsec:theory_validation}

\subsubsection{Existence of an Optimum}
We consider a quasi-stationary network condition $\omega$ (e.g., cross-traffic intensity, queueing state, end-host contention) over a short monitoring interval (MI). For any choice of concurrency $cc$ and parallelism $p$ within predefined safe bounds, the transfer produces measurable outcomes: throughput $T(cc,p;\omega)$, loss $L(cc,p;\omega)$, and energy $E(cc,p;\omega)$. We denote the total number of TCP streams by
\begin{equation}
n \triangleq cc\cdot p,\qquad (cc,p)\in\mathcal{A},
\end{equation}
where $\mathcal{A}$ is a finite feasible set (bounded by $cc_{\min}\!\le cc\!\le cc_{\max}$ and $p_{\min}\!\le p\!\le p_{\max}$).

For a fixed $\omega$, the instantaneous utility (either fairness \& efficiency or throughput/energy) is therefore a function over a \emph{finite} action set:
\begin{equation}
(cc^*(\omega),p^*(\omega))\in \arg\max_{(cc,p)\in\mathcal{A}} U(cc,p;\omega).
\label{eq:static_opt_ccp}
\end{equation}
Hence, the static optimization is solvable. The practical difficulty is not existence, but that $U(\cdot;\omega)$ is unknown a priori and the underlying condition $\omega$ changes during the transfer, so the maximizer drifts over time.

\subsubsection{Why a Dominant Optimum is Expected}
For loss-based TCP on wide-area paths, increasing the number of streams typically yields (i) diminishing throughput gains near the bottleneck capacity and (ii) sharply increasing congestion signals (loss and RTT inflation) once the bottleneck is persistently overfilled. Combined with stream-management overhead, this produces a ``knee'': below the knee, adding streams helps; beyond it, congestion and overhead dominate.

Formally, for a quasi-stationary $\omega$, the following mild behaviors are commonly observed:
\begin{itemize}
    \item $T(cc,p;\omega)$ increases with diminishing returns as the bottleneck saturates,
    \item $L(cc,p;\omega)$ stays low initially and rises rapidly once persistent queueing begins,
    \item overhead increases with the stream budget $n=cc\cdot p$ (e.g., via the $K^{n}$ term in Eq.~\eqref{eq:utility_fe}).
\end{itemize}
These effects yield a utility landscape with a \emph{dominant peak} around the knee, i.e., a unimodal (quasi-concave) structure over the feasible $(cc,p)$ grid for a fixed $\omega$.

\begin{figure}[t]
   \includegraphics[width=0.40\textwidth]{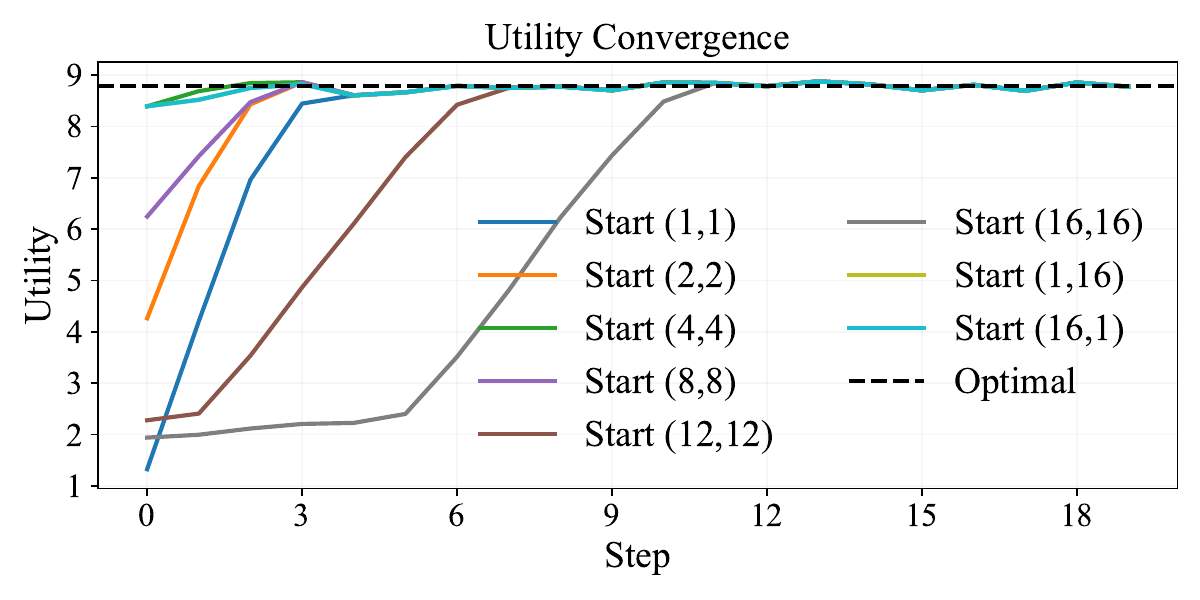}
   \caption{Local-search trajectories. Local updates from diverse initializations converge to the same maximizer, suggesting small bounded steps are sufficient to reach (and track) the optimum. }
\vspace{-3mm}
\label{fig:convergence}
\end{figure}

Figure~\ref{fig:utility_surface} abd ~\ref{fig:utility_surface_te} visualizes exactly this: the measured utility over $(cc,p)$ exhibits one clear maximizer (marked in the plot) and decreases smoothly away from that point, indicating that aggressively increasing streams eventually reduces utility due to congestion and overhead. Figure~\ref{fig:convergence} further supports \emph{trackability}: starting from diverse initial configurations, a local update rule converges to the same maximizer within a small number of steps. This behavior motivates our bounded, small-step action design ($\pm1,\pm2$), which is sufficient to reach (and later track) the moving optimum as $\omega_t$ changes.




\vspace{-4mm}
\subsection{RL Formulation (POMDP)}
\label{subsec:pomdp_and_objective}

Because the true network condition $\omega_t$ is not directly observable at end hosts, we model tuning as a Partially Observable Markov Decision Process (POMDP).
At each MI $t$, the agent observes an end-host feature history $s_t$, selects an action $a_t$ that updates $(cc_t,p_t)$, and receives reward $r_t$ defined in Eq.~\eqref{eq:reward_def}.
The RL objective is:
\begin{equation}
\max_{\theta}\; \mathbb{E}_{\pi_\theta}\!\left[\sum_{t=t_s}^{t_f}\gamma^{t-t_s} r_t\right],
\qquad \gamma\in(0,1],
\label{eq:drl_objective}
\end{equation}
where $\pi_\theta(a_t\mid s_t)$ is the learned policy parameterized by $\theta$.
Intuitively, the agent learns to increase $n_t$ when the path is underutilized and to decrease $n_t$ when loss/RTT trends indicate rising congestion, thereby tracking the moving optimum implied by the unimodal utility geometry.

\vspace{-4mm}
\subsection{State, Action and Reward Design}
\label{subsec:reward_formulation}



\subsubsection{RL State Space}
\label{subsec:state}

The agent must infer congestion from end-host signals, so the state uses \emph{stable indicators} rather than the optimization targets (throughput/energy). At each monitoring interval (MI) $t$, we form:
\begin{equation}
x_t = \{plr_t,\ rtt\_gradient_t,\ rtt\_ratio_t,\ cc_t,\ p_t\},
\end{equation}
where $plr_t$ is packet loss rate, $rtt\_gradient_t$ captures RTT trend, and $rtt\_ratio_t$ normalizes current RTT by the best (minimum) RTT observed so far~\cite{nathan-jay2019}. We include $(cc_t,p_t)$ so the policy can associate parameter choices with subsequent congestion signals. To mitigate partial observability and measurement noise, the agent observes a short history window:
\begin{equation}
s_t = (x_{t-h+1},\ldots,x_t),
\end{equation}
which allows the policy to detect trends (e.g., rising RTT and loss) and react before severe congestion.

\subsubsection{Action Space and Parameter Constraints}
\label{subsec:actions}

All evaluated RL methods (PPO~\cite{schulman2017proximal}, R\_PPO~\cite{kapturowski2019recurrent}, DDPG~\cite{lillicrap2016continuous}, DQN~\cite{Mnih2015}, DRQN~\cite{hausknecht2015deep}) use the same discrete action set of five \emph{joint} updates to $(cc,p)$. At MI $t$, the agent selects $a_t\in\{0,1,2,3,4\}$:
\begin{align*}
a_t=0 &: (cc_{t+1},p_{t+1})=(cc_t,p_t)\\
a_t=1 &: (cc_{t+1},p_{t+1})=(cc_t+1,p_t+1)\\
a_t=2 &: (cc_{t+1},p_{t+1})=(cc_t-1,p_t-1)\\
a_t=3 &: (cc_{t+1},p_{t+1})=(cc_t+2,p_t+2)\\
a_t=4 &: (cc_{t+1},p_{t+1})=(cc_t-2,p_t-2).
\end{align*}
We clip updates to enforce operational bounds:
\begin{equation}
cc_{\min}\le cc_t\le cc_{\max},\qquad p_{\min}\le p_t\le p_{\max},
\end{equation}
and initialize from a moderate starting point (e.g., $(cc_0,p_0)=(4,4)$) to avoid extreme exploration early in an episode. Using a shared, bounded action space ensures controlled exploration (stability on real networks) and enables fair comparison across RL algorithms.

\subsubsection{Reward Computation and Smoothing}
We compute $r_t$ using Eq.~\eqref{eq:reward_def}. To reduce measurement noise, we optionally use a short moving average:
\begin{equation}
\tilde{r}_t = \frac{1}{h}\sum_{i=t-h+1}^{t} r_i,
\label{eq:reward_smooth}
\end{equation}
and train on $\tilde{r}_t$. 

\begin{figure}[t]
  \centering
  \includegraphics[width=0.45\textwidth]{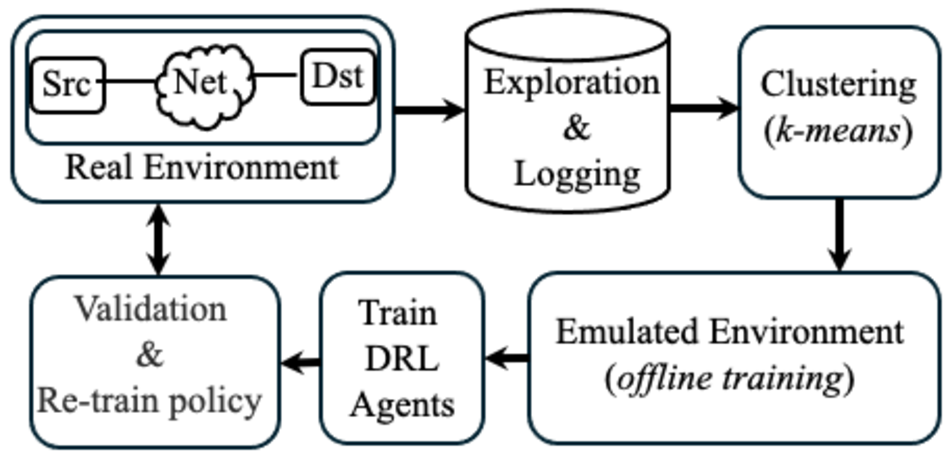}
  \caption{\small Offline--online workflow: exploratory data collection in the real environment, clustering of transitions, offline training in the emulator, and periodic validation / re-training with new logs.}
  \label{fig:blockdiagram1}
\end{figure}





\vspace{-3mm}
\subsection{Offline--Online Training via an Emulated Environment}
\label{sec:emulated-training-env}

Training an RL controller purely online is expensive because early exploration produces many suboptimal transfers. To reduce time and energy overhead, we train primarily \emph{offline} using an emulator built from previously recorded transitions, and then validate (and optionally fine-tune) the policy online. Figure~\ref{fig:blockdiagram1} summarizes this workflow.

\paragraph{Step 1: High-exploration data collection (real path)}
We first run the agent on the real network under a high-exploration policy and log per-MI (or per-second) measurements, including throughput, loss rate ($plr$), RTT, energy, and the current $(cc,p)$. Each log entry is labeled with the reward/utility value produced by the chosen objective. Conceptually, this yields a dataset of transitions
\begin{equation}
\mathcal{D}=\{(s_t,a_t,r_t,s_{t+1})\},
\end{equation}
where $s_t$ is the observed state (Section~\ref{subsec:state}), $a_t$ is the discrete action (Section~\ref{subsec:actions}), and $r_t$ is the reward.

\paragraph{Step 2: Clustering transitions into ``network scenarios''}
We map each MI to a compact feature vector (e.g., $x_t=[plr_t,\,rtt\_gradient_t,\,rtt\_ratio_t,\,cc_t,\,p_t]$) and cluster transitions using $k$-means with $k=40$ clusters~\cite{lloyd1982least}. Each cluster groups similar state--action contexts and their observed outcomes, and its centroid represents a
recurring ''network scenario,'' enabling a simplified lookup when simulating new training episodes.
\begin{figure}[t]
    \centering
    \includegraphics[width=0.40\textwidth]{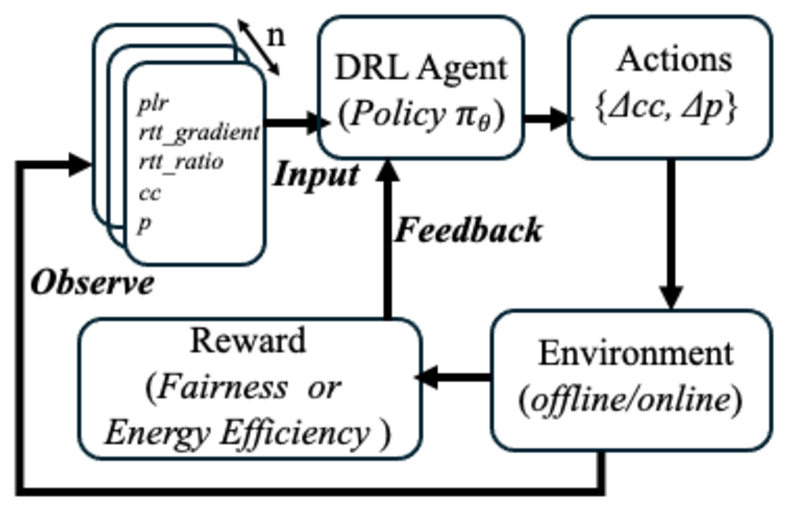}
    \caption{\small RL loop for transfer tuning: the agent observes state features (history of $plr$, RTT trends, and current $(cc,p)$), selects an action ($\Delta cc,\Delta p$), and receives reward consistent with the selected objective.}
    \label{fig:blockdiagram-2}
\end{figure}

\paragraph{Step 3: Lookup-based emulation for offline training}
The emulator behaves like an environment that returns plausible next states and rewards without executing a real transfer:
\begin{enumerate}
    \item \emph{Initialize:} sample an initial state $s_1$ from $\mathcal{D}$.
    \item \emph{Step:} given $(s_t,a_t)$, select the nearest cluster (or clusters) and uniformly sample one transition from that cluster to produce $(s_{t+1},r_t)$.
\end{enumerate}
Sampling within a cluster introduces variability (multiple plausible outcomes) and reduces overfitting compared to a deterministic lookup. Figure~\ref{fig:comparisionAlgos} shows that trends learned offline in the emulator transfer to real deployments: algorithms that achieve higher throughput (or lower energy) in emulation exhibit consistent behavior during real transfers, supporting the emulator’s fidelity for offline training.

\paragraph{Step 4: Online validation and refresh.}
After the policy converges offline, we deploy it on the real environment for validation. If performance degrades due to a new network regime, we can collect additional transitions, re-cluster, and refresh the emulator or retune the policy in real world environment.

\vspace{-5mm}
\subsection{Evaluated RL Algorithms and Training Procedure}
\label{sec:drl_algorithms}

We evaluate representative on-policy and off-policy RL methods for tuning $(cc,p)$:
DQN~\cite{Mnih2015}, DRQN~\cite{hausknecht2015deep}, PPO~\cite{schulman2017proximal}, R\_PPO~\cite{rpposb3} , and DDPG~\cite{lillicrap2016continuous}. We include recurrent variants (DRQN, R\_PPO) because end hosts only observe indirect congestion signals; recurrence provides memory that improves decisions under partial observability. Figure~\ref{fig:blockdiagram-2} shows the closed loop: the agent observes the recent state, selects an action to update $(cc,p)$, and receives a reward defined by the selected objective. 
\begin{algorithm}[h]
\caption{\texttt{SPARTA} Training Procedure}
\scriptsize
\label{alg:general-training}
\begin{algorithmic}[1]
\REQUIRE 
  \textbf{Models:} A chosen RL model from \{\textit{DQN}, \textit{DRQN}, \textit{PPO}, \textit{R\_PPO}, \textit{DDPG}\}; \\
  \textbf{Environment:} Provides states $s$, actions $a$, and rewards $r$ \\
  \textbf{Hyperparameters:} Learning rate, discount factor $\gamma$, exploration rate, etc.
\ENSURE 
  A trained policy $\pi_\theta(a \mid s)$ (and possibly a value function $V_\psi(s)$ or $Q_\psi(s,a)$) 
  for optimizing concurrency ($cc$) and parallelism ($p$)

\STATE \textbf{Initialize} model parameters (e.g., $\theta$, $\psi$ for actor--critic), set replay buffer $\mathcal{D}$ empty for off-policy methods
\STATE Define maximum number of training episodes $N$ and per-episode length $T$

\FOR{episode $= 1$ \TO $N$}
  \STATE Reset environment and obtain initial state $s_1$
  \FOR{$t = 1$ \TO $T$}
    \STATE \textbf{Observe current state} $s_t$ (e.g., $plr$, $rtt\_gradient$, $rtt\_ratio$, $cc$, $p$)
    \STATE \textbf{Select action} $a_t$ according to the RL method:
      \begin{itemize}
         \item \textbf{On-policy} (\textit{PPO}, \textit{R\_PPO}): sample $a_t \sim \pi_\theta(a\mid s_t)$
         \item \textbf{Off-policy} (\textit{DQN}, \textit{DRQN}, \textit{DDPG}):
         \begin{itemize}
           \item \textbf{DQN/DRQN:} $\epsilon$-greedy over $Q_\psi(s_t,a)$
           \item \textbf{DDPG:} $a_t = \pi_\theta(s_t) + \text{noise}$
         \end{itemize}
      \end{itemize}
    \STATE Execute $a_t$ in the environment to get new state $s_{t+1}$ and reward $r_t$
    \STATE \textbf{Store transition} $(s_t, a_t, r_t, s_{t+1})$:
      \begin{itemize}
        \item For \textbf{off-policy} methods, place in replay buffer $\mathcal{D}$
        \item For \textbf{on-policy} methods, store in the current trajectory
      \end{itemize}
    \STATE \textbf{Update model parameters} (frequency-dependent):
      \begin{itemize}
        \item \textbf{DQN/DRQN:} sample mini-batches from $\mathcal{D}$ to minimize:
          \[
            \bigl(Q_\psi(s,a) - [r + \gamma \max_{a'}Q_\psi(s_{t+1},a')]\bigr)^2
          \]
        \item \textbf{DDPG:} sample mini-batches from $\mathcal{D}$ to update critic:
          \[
            y_t = r + \gamma\,Q_\psi\bigl(s_{t+1}, \pi_\theta(s_{t+1})\bigr)
          \]
          and use policy gradient to update the actor
        \item \textbf{PPO/R\_PPO:} after collecting a rollout, optimize clipped surrogate objective:
          \[
            \mathcal{L}_{\pi} = \mathbb{E}_t \Bigl[\min\bigl(r_t(\theta)\cdot A_t,\, \mathrm{clip}(r_t(\theta), 1-\epsilon, 1+\epsilon)\cdot A_t\bigr)\Bigr]
          \]
      \end{itemize}
    \IF{end of episode or environment termination}
      \STATE \textbf{break}
    \ENDIF
  \ENDFOR
\ENDFOR
\RETURN policy $\pi_\theta$ (and $Q_\psi$ or $V_\psi$ if applicable).
\end{algorithmic}
\end{algorithm}

\noindent\textbf{Algorithm overview.}
Algorithm~\ref{alg:general-training} follows a standard RL loop: initialize the model and (if applicable) a replay buffer; repeatedly interact with the environment by observing $s_t$, choosing $a_t$, receiving $(r_t,s_{t+1})$, storing experience, and updating the model. On-policy methods (PPO, R\_PPO) update from recent rollouts of the current policy, while off-policy methods (DQN, DRQN, DDPG) reuse stored transitions via a replay buffer for sample-efficient learning. The output is a trained policy that selects $(cc,p)$ online from end-host observations.

\begin{table*}[t]
\centering
\begin{adjustbox}{width=1.95\columnwidth}
\begin{tabular}{@{}lccccccccc@{}}
\toprule
Method & \shortstack{Offline Training \\ Time (min) $\downarrow$} & \shortstack{Steps in\\  Converging $\downarrow$} & \shortstack{Training Avg \\ CPU\% $\downarrow$} & \shortstack{Training Avg \\ GPU\% $\downarrow$} & \shortstack{Training Avg \\ Memory\% $\downarrow$} & \shortstack{Training \\ Energy(KJ) $\downarrow$} & \shortstack{Inference \\ Time (ms) $\downarrow$} & \shortstack{Model Inference\\ Energy (J) $\downarrow$} & \shortstack{Energy During \\ Online Tuning (KJ)$\downarrow$} \\
\midrule
\rowcolor{gray!10} DQN & \best{30} & \best{300K} & 26.62 & \best{10.8} & \best{16.74} & \best{131} & 0.57 & 0.098 & 4.3 \\
PPO & 36 & \worst{600K} & 26.31 & 11.71 & 16.75 & 158 & \worst{0.74} & \best{0.088} & \best{2.6} \\
\rowcolor{gray!10} DDPG & 59 & 320K & \best{26.11} & \worst{25.59} & \worst{20.74} & \worst{294} & 0.58 & 0.091 & \worst{9.18} \\
R\_PPO & 48.5 & 550K & \worst{26.93} & 15.41 & 19.27 & 221 & 0.66 & \worst{0.094} & 4.01 \\
\rowcolor{gray!10} DRQN & \worst{94} & 400K & 26.67 & 12.67 & 17.66 & 214 & \best{0.21} & \best{0.088} & 5.35 \\
\bottomrule
\end{tabular}
\end{adjustbox}
\caption{Comparison of RL algorithms trained for 100000 steps with associated training metrics.}
\label{table:comparison}
\end{table*}

\vspace{-5mm}
\subsection{Comparison of RL Solutions}
\label{sec:rl-evaluation}

We compare five RL algorithms (DQN, PPO, DDPG, R\_PPO, DRQN) under the throughput-focused energy objective (T/E). We first describe the offline training data and emulator, then summarize training/inference costs (Table~\ref{table:comparison}), and finally compare transfer throughput and energy in simulation and on a real wide-area environment (Figure~\ref{fig:comparisionAlgos}).

\subsubsection{Offline Training Setup}
We collect state--action transitions on a real testbed (Chameleon Cloud, TACC$\rightarrow$UC) by sweeping a wide range of $(cc,p)$. Each per-MI (or per-second) record contains the state features ($plr$, $rtt\_gradient$, $rtt\_ratio$, $cc$, $p$), the action taken, and the resulting throughput and energy. We then train each RL method offline using the clustering-based emulator (Section~\ref{sec:emulated-training-env}), which avoids repeated low-efficiency transfers during exploration.

\subsubsection{Training and Inference Cost (Table~\ref{table:comparison})}
Table~\ref{table:comparison} reports three types of overhead:

\begin{itemize}
    \item \textbf{Offline training cost.} DQN converges fastest and with the lowest training energy (30\,min, 300K steps, 131\,KJ). DRQN is slowest (94\,min), and DDPG is the most energy-intensive to train (294\,KJ), consistent with heavier actor--critic updates and higher GPU utilization.
    \item \textbf{Inference cost.} Per-step inference is sub-millisecond for all methods. DRQN has the lowest latency (0.21\,ms), while PPO is the slowest (0.74\,ms). Inference energy is small across the board (0.088--0.098\,J/step).
    \item \textbf{Online tuning cost.} When adapting to a new environment, PPO requires the least additional tuning energy (2.6\,KJ), whereas DDPG requires the most (9.18\,KJ).
\end{itemize}

Overall, Table~\ref{table:comparison} quantifies the compute/energy overhead of each learner and motivates the accuracy--overhead trade-offs discussed next.

\subsubsection{Transfer Throughput and Energy}
\begin{figure*}[t]
\centering
  \begin{subfigure}[b]{0.90\columnwidth}
    \includegraphics[width=\linewidth]{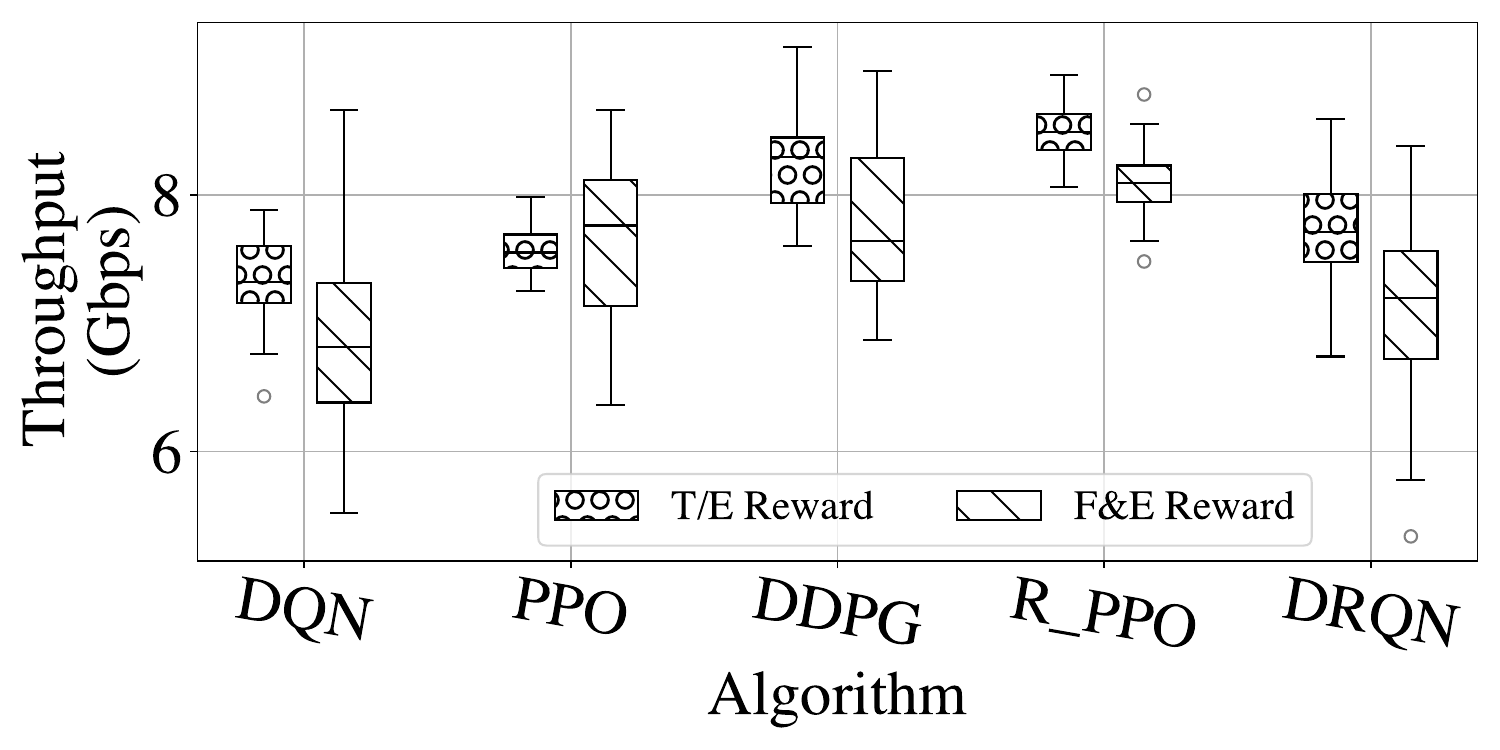}
    \caption{Throughput in simulation}
    \label{fig:simu_throughput}
  \end{subfigure}
  \begin{subfigure}[b]{0.90\columnwidth}
    \includegraphics[width=\linewidth]{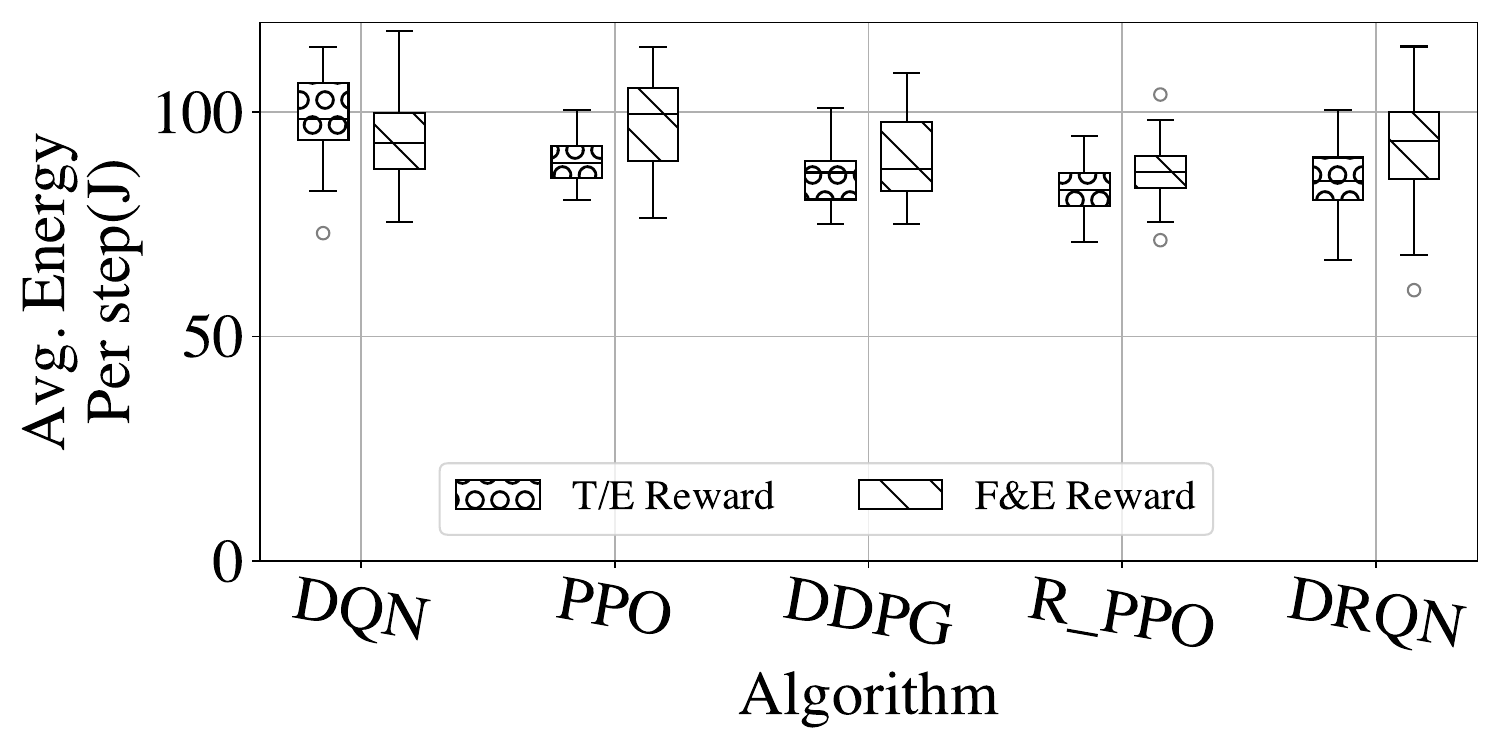}
    \caption{Energy in simulation}
    \label{fig:simu_energy}
  \end{subfigure}
\\
  \begin{subfigure}[b]{0.90\columnwidth}
    \includegraphics[width=\linewidth]{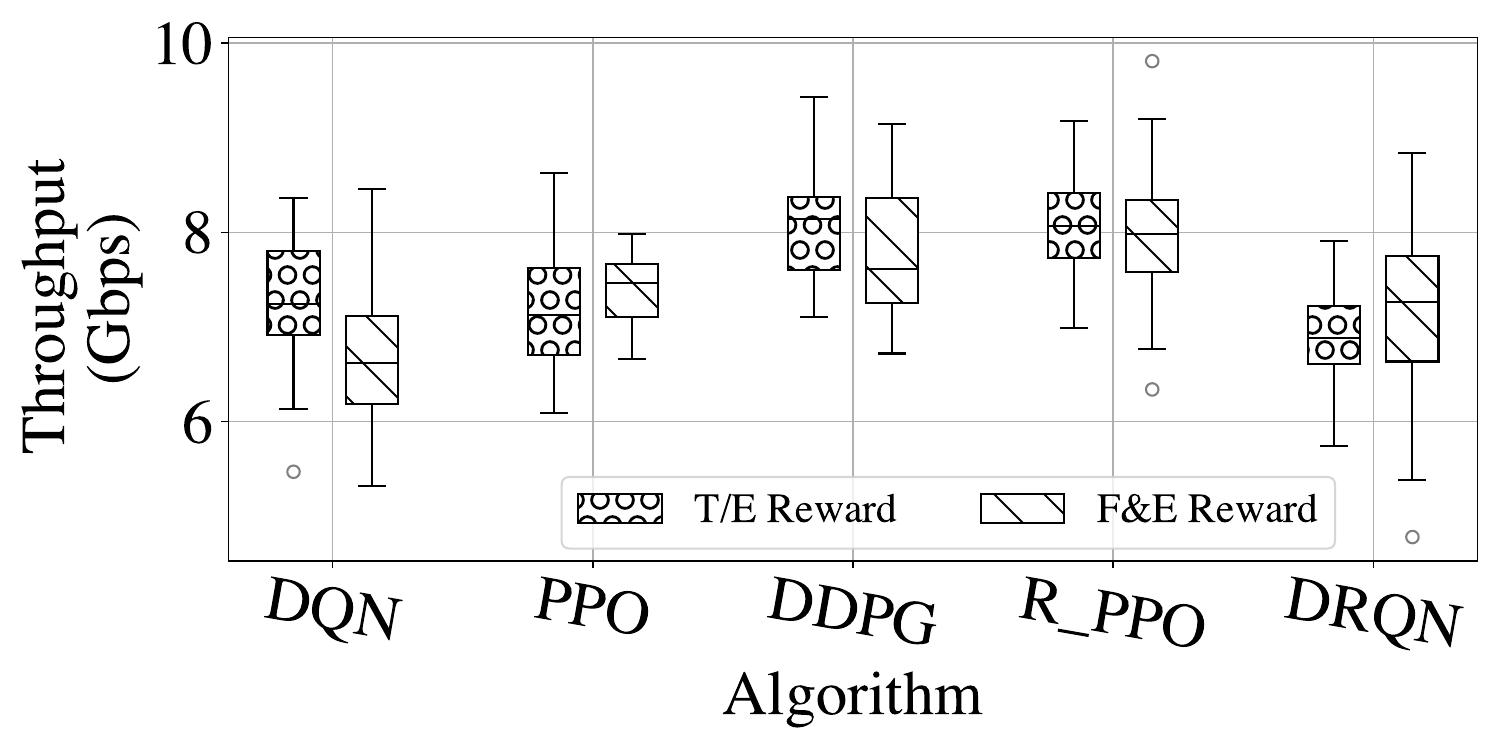}
    \caption{Throughput in real transfer}
    \label{fig:real_throughput}
  \end{subfigure}
  \begin{subfigure}[b]{0.90\columnwidth}
    \includegraphics[width=\linewidth]{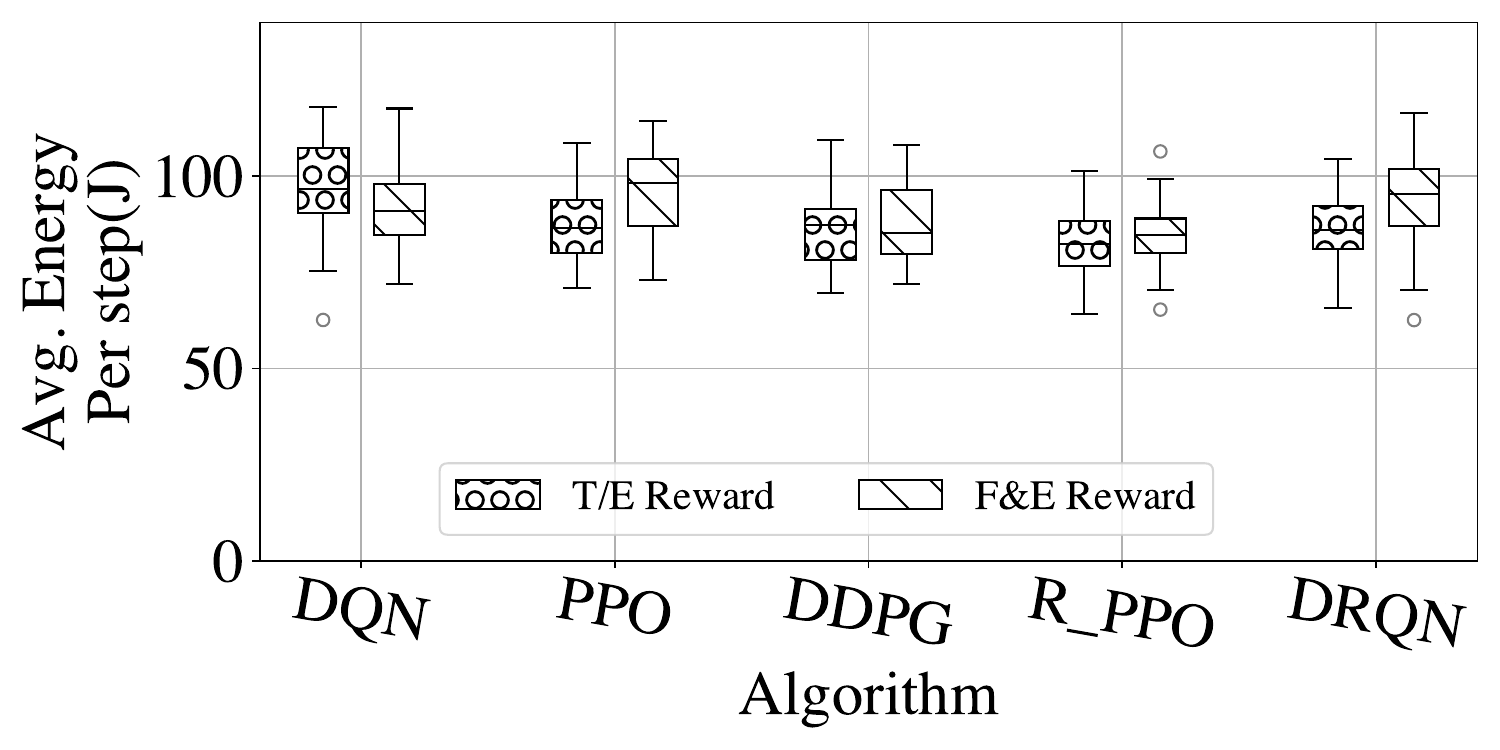}
    \caption{Energy in real transfer}
    \label{fig:real_energy}
  \end{subfigure}
  \caption{\small Comparison of RL algorithms trained offline and evaluated in simulation and real transfers.}
  \label{fig:comparisionAlgos}
\end{figure*}

Figure~\ref{fig:comparisionAlgos} compares achieved throughput and transfer energy in both simulation (top) and real transfers (bottom). In simulation, DDPG tends to push higher throughput but with higher energy, while DQN and PPO deliver moderate throughput with lower energy. Recurrent policies (R\_PPO, DRQN) are generally more stable under fluctuating conditions because memory helps disambiguate partially observed congestion.

On the real environment, the same trends largely hold with increased variance: DDPG remains aggressive (higher throughput, higher energy), DQN/PPO are more conservative, and R\_PPO typically sustains steadier throughput while controlling transfer energy by adapting $(cc,p)$ quickly. Performance gaps between simulation and real runs indicate incomplete generalization from offline data to unseen network regimes, motivating online tuning when deployment conditions change.

\begin{figure}[t]
  \centering
  \includegraphics[width=0.95\columnwidth]{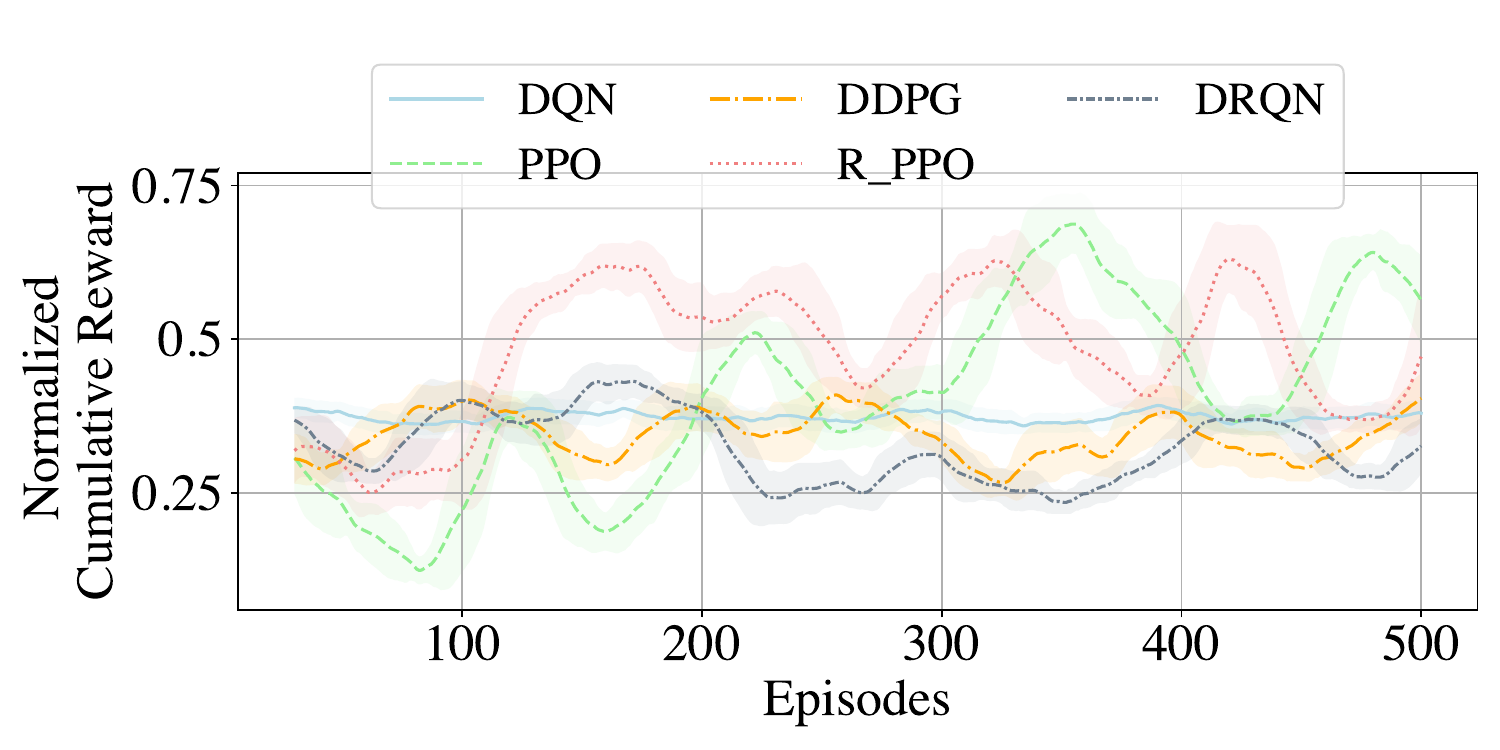}
  \caption{\small Cumulative reward progression over 500 episodes for agents trained on one environment and tuned on another. }
  \label{fig:retrain}
\end{figure}

\subsubsection{Online Tuning Performance and Final Selection}
Figure~\ref{fig:retrain} shows the cumulative reward over 500 episodes when agents—trained on Chameleon nodes for throughput focused energy objective (T/E)—are deployed to CloudLab nodes. Early on, all algorithms see a dip in performance due to the new environment’s distinct RTT patterns and congestion levels. Over time, most agents adapt, but their final reward levels vary, indicating different degrees of generalization:

\begin{itemize}
    \item DQN and DDPG initially underperform and display larger fluctuations. Although they improve with more episodes, their cumulative rewards remain lower.
    \item PPO adapts more smoothly, maintaining a higher reward than DQN and DDPG. Its on-policy updates help it generalize, but it takes longer to match its prior performance.
    \item R\_PPO stands out for quickly reaching a higher reward plateau, reflecting both on-policy stability and recurrent memory that better handles the new network’s partial observability. By episode 200, R\_PPO has already surpassed the other algorithms, indicating stronger generalization.
\end{itemize}

Overall, these tuning trajectories highlight that agents not fully tailored to the new environment can struggle to achieve higher rewards, whereas PPO- and especially R\_PPO-based policies exhibit more robust adaptation and faster convergence under different network conditions.

\subsubsection{Why R\_PPO for \texttt{SPARTA}? Training and Inference Cost Analysis}
\label{subsec:why_rppo}

Table~\ref{table:comparison} shows that R\_PPO has higher \emph{one-time} training cost and slightly higher per-step inference overhead than some baselines (e.g., DQN, PPO). However, Figures~\ref{fig:comparisionAlgos} and~\ref{fig:retrain} show that R\_PPO achieves better \emph{transfer-level} behavior (higher throughput and/or lower data-plane energy). Since \texttt{SPARTA} is trained once and then used repeatedly, the relevant question is: \emph{after how many transfers does R\_PPO's lower transfer energy outweigh its higher training cost?}

We quantify this trade-off by accounting for both training energy and per-transfer energy. Let:
\begin{itemize}
    \item \(E_{\text{train}}(M)\): energy to train model \(M\) (J),
    \item \(E_{\text{inf}}(M)\): inference energy per monitoring interval (MI) step (J/step),
    \item \(E_{\text{data}}(M)\): data-plane energy per MI step during transfer (J/step),
    \item \(S\): number of MI steps per transfer,
    \item \(T\): number of transfers executed using model \(M\).
\end{itemize}
The total energy over \(T\) transfers is
\begin{align}
\text{TotalCost}(M) = E_{\text{train}}(M) + T \times S \times \Bigl[ E_{\text{inf}}(M) + E_{\text{data}}(M) \Bigr],
\end{align}
and the average energy per transfer is
\begin{align}
\scriptsize
\bar{O}(M) = \frac{\text{TotalCost}(M)}{T} = \frac{E_{\text{train}}(M)}{T} + S \times \Bigl[ E_{\text{inf}}(M) + E_{\text{data}}(M) \Bigr].
\end{align}
As \(T\) grows, the amortized term \(E_{\text{train}}(M)/T\) becomes small, and the dominant cost is the per-transfer data-plane energy.

To find the break-even point where R\_PPO becomes more energy-efficient than an alternative model \(M'\), we solve:
\begin{equation}
\begin{aligned}
E_{\text{train}}(\text{R\_PPO}) + T \,S \,\Bigl[ E_{\text{inf}}(\text{R\_PPO}) + E_{\text{data}}(\text{R\_PPO}) \Bigr] \\
<\; E_{\text{train}}(M') + T \,S \,\Bigl[ E_{\text{inf}}(M') + E_{\text{data}}(M') \Bigr].
\end{aligned}
\end{equation}
Rearranging yields:
\begin{align}
\scriptsize
T > \frac{E_{\text{train}}(\text{R\_PPO}) - E_{\text{train}}(M')}
{S \Bigl[ \bigl(E_{\text{inf}}(M') + E_{\text{data}}(M')\bigr) - \bigl(E_{\text{inf}}(\text{R\_PPO}) + E_{\text{data}}(\text{R\_PPO})\bigr) \Bigr]}.
\end{align}

\paragraph{Example (vs.\ DQN).}
Assume \(S = 1250\) steps per transfer (e.g., transferring 1000 files of 1\,GB each). Using Table~\ref{table:comparison} and Figure~\ref{fig:real_energy}:
\begin{itemize}
    \item R\_PPO: \(E_{\text{train}}=221\,K\,\text{J}\), \(E_{\text{inf}}\approx 0.094\,\text{J/step}\), \(E_{\text{data}}\approx 90\,\text{J/step}\).
    \item DQN: \(E_{\text{train}}=131\,K\,\text{J}\), \(E_{\text{inf}}\approx 0.098\,\text{J/step}\), \(E_{\text{data}}\approx 110\,\text{J/step}\).
\end{itemize}
Substituting:
\[
\begin{aligned}
T &> \frac{221\,K - 131\,K}{1250 \times \Bigl[ (0.098 + 110) - (0.094 + 90) \Bigr]}
\approx 3.6.
\end{aligned}
\]
Thus, after roughly 4 transfers, the training-energy premium of R\_PPO is recovered by its lower per-transfer energy. Since \texttt{SPARTA} is intended for repeated use (often hundreds to thousands of transfers), the one-time training overhead is quickly amortized, which justifies selecting R\_PPO despite its higher upfront cost.

\vspace{-5mm}
\subsection{\textit{SPARTA} Models}
\label{subsec:sparta_models}

We train two \texttt{SPARTA} variants. Both use the same R\_PPO policy architecture and state/action design; they differ only in the reward (Section~\ref{subsec:reward_formulation}):

\begin{itemize}
    \item {\texttt{SPARTA-FE}} (Fairness \& Efficiency): uses the F\&E reward (Eqs.~\ref{eq:reward_def}--\ref{eq:reward_smooth}) to maximize throughput while penalizing congestion (via packet loss), which promotes fair sharing and reduces wasted work from retransmissions/queueing.
    \item {\texttt{SPARTA-T}} (Throughput-focused Energy): uses the T/E reward (Eqs.~\ref{eq:reward_def}--\ref{eq:reward_smooth}) to maximize throughput per unit energy, explicitly favoring energy-efficient parameter settings.
\end{itemize}

This separation lets operators choose the objective that matches the deployment goal (fairness/congestion control vs.\ energy efficiency) without changing the learning algorithm.

\section{Evaluation}
\label{sec:evaluation}

We assess the performance of \texttt{SPARTA} and compare it with state-of-the-art solutions on multiple testbeds (e.g., CloudLab, Chameleon Cloud, and FABRIC) with different network configurations, measuring throughput, energy consumption, and fairness. In each trial, we send \(1000 \times 1\text{GB}\) files from a dedicated sender to a receiver. We repeat the entire transfer procedure five times in sequence, collecting performance and resource usage metrics for each run. Our results are shown as distributions aggregated from these five trials.
\begin{figure*}[t]
  \centering
  \begin{subfigure}[b]{0.32\textwidth} 
    \includegraphics[width=\textwidth]{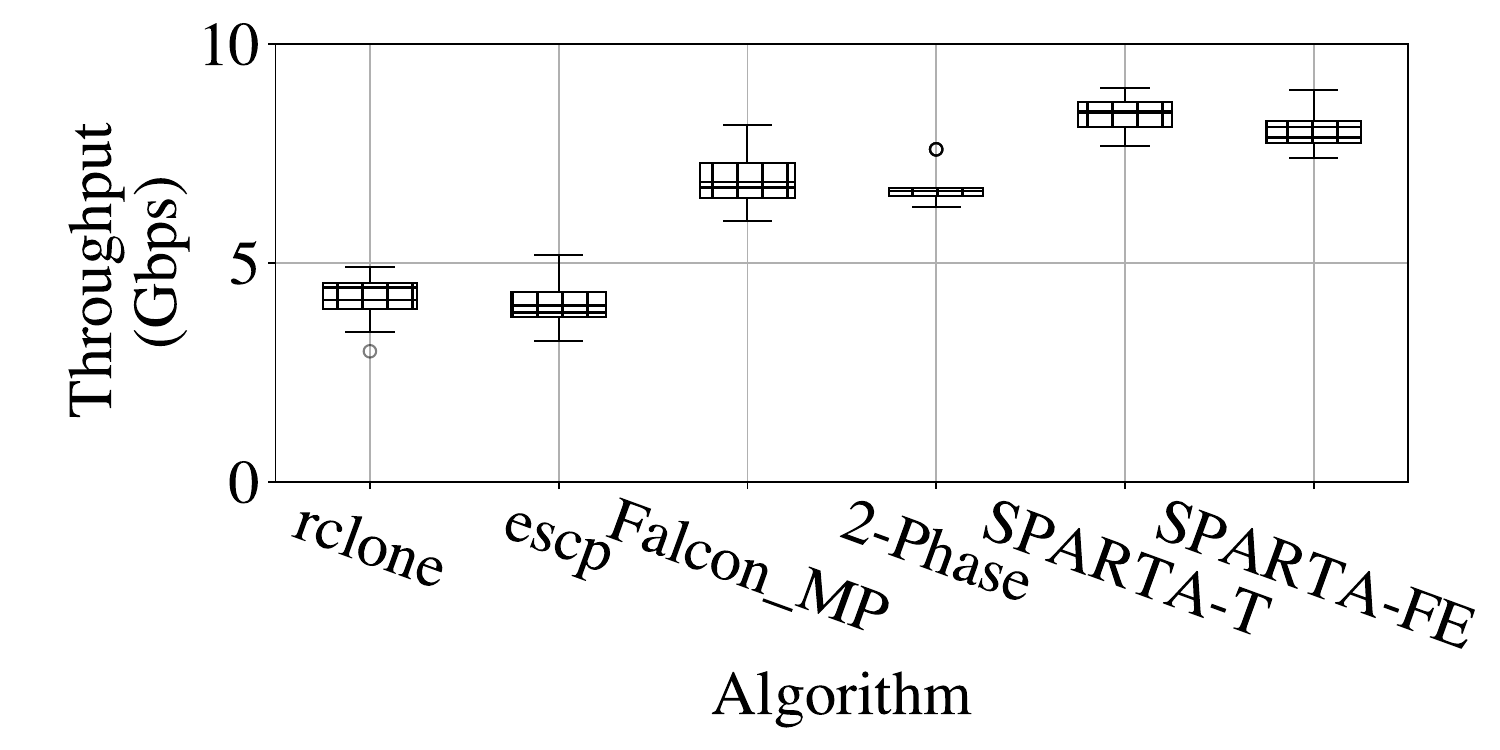}
    \caption{\small Throughput in Chameleon}
    \label{fig:cham_th}
  \end{subfigure}
  \hfill
  \begin{subfigure}[b]{0.32\textwidth}
    \includegraphics[width=\textwidth]{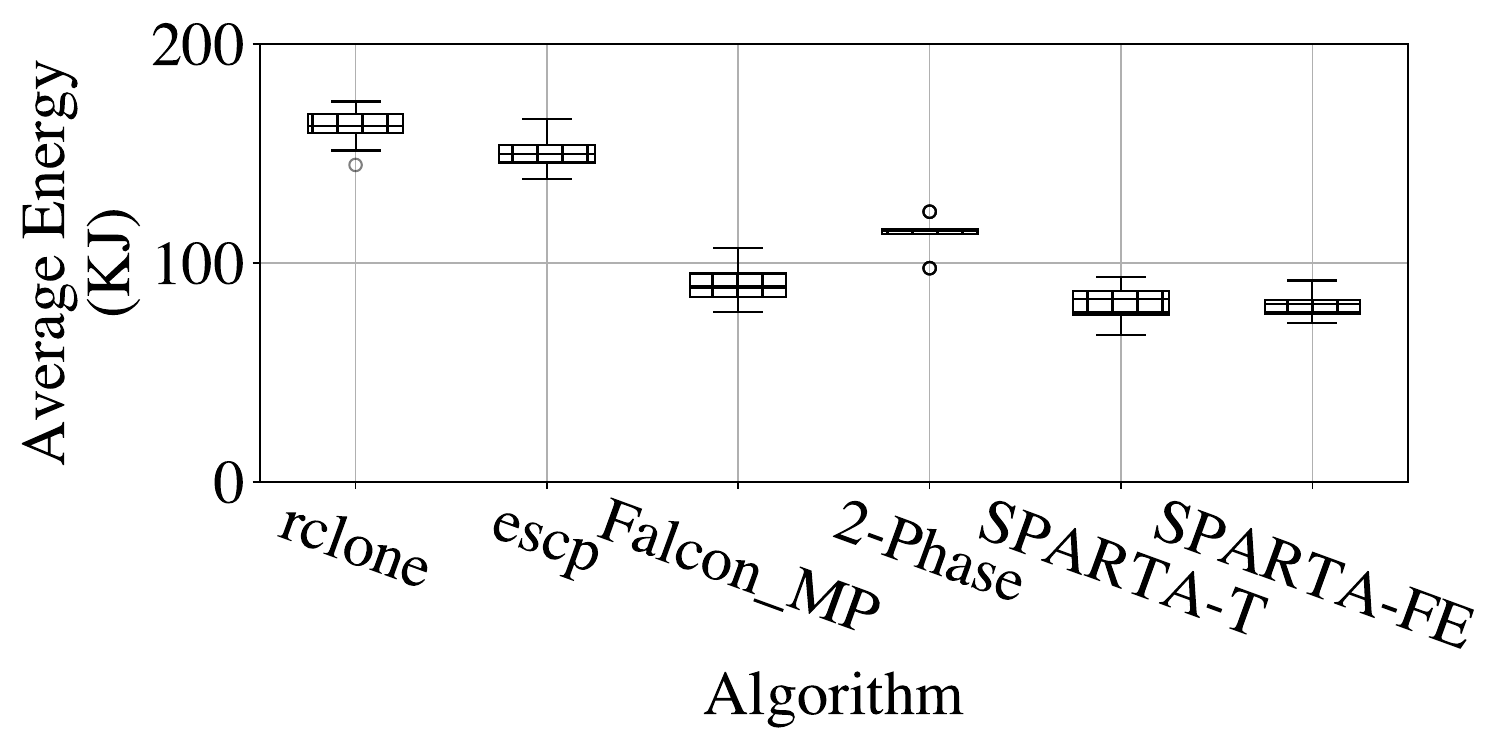}
    \caption{\small Energy in Chameleon}
    \label{fig:cham_en}
  \end{subfigure}
  \hfill
  \begin{subfigure}[b]{0.32\textwidth}
    \includegraphics[width=\textwidth]{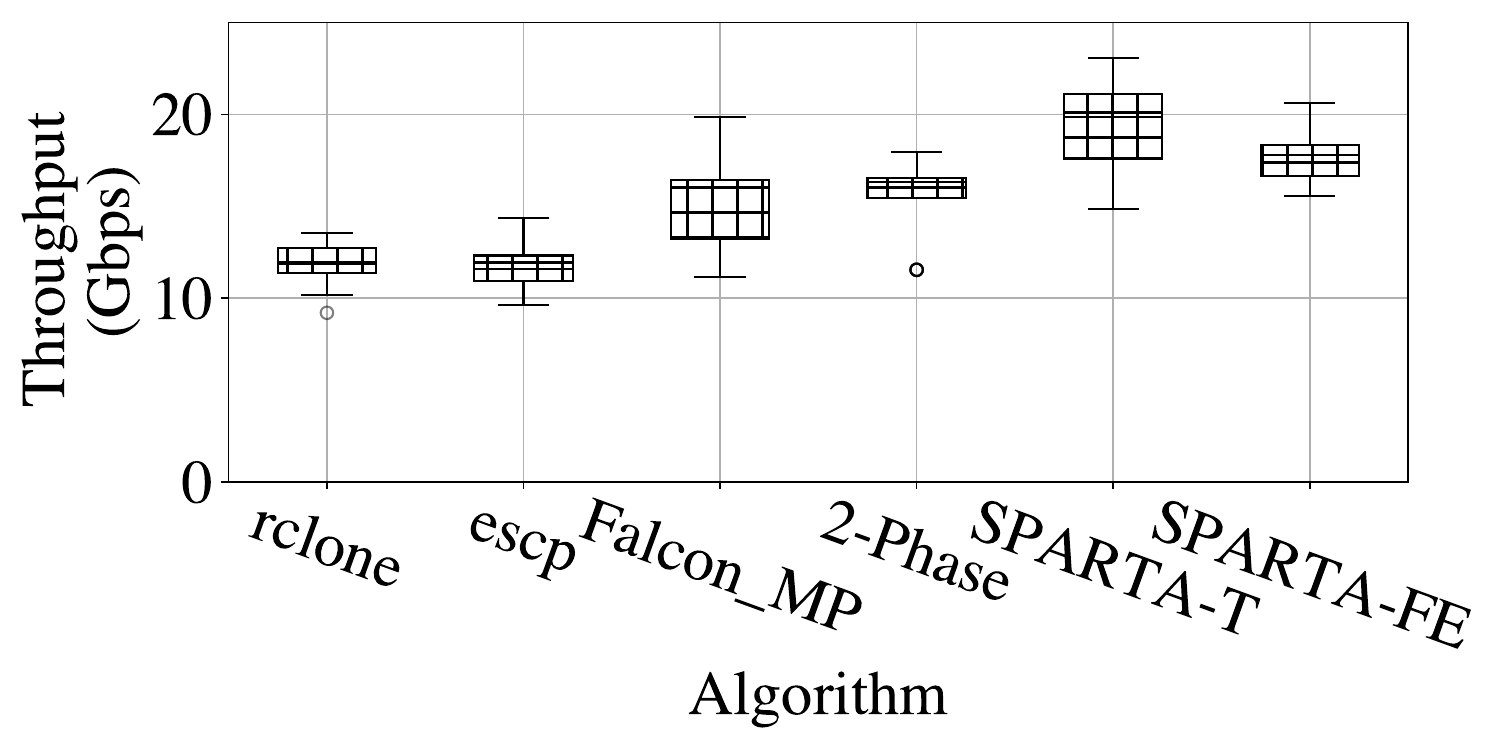}
    \caption{\small Throughput in Cloudlab}
    \label{fig:cloudlab_th}
  \end{subfigure}
  
  \vspace{1mm} 
   \hspace*{0.10\columnwidth} 
  \begin{subfigure}[b]{0.32\textwidth} 
    \includegraphics[width=\textwidth]{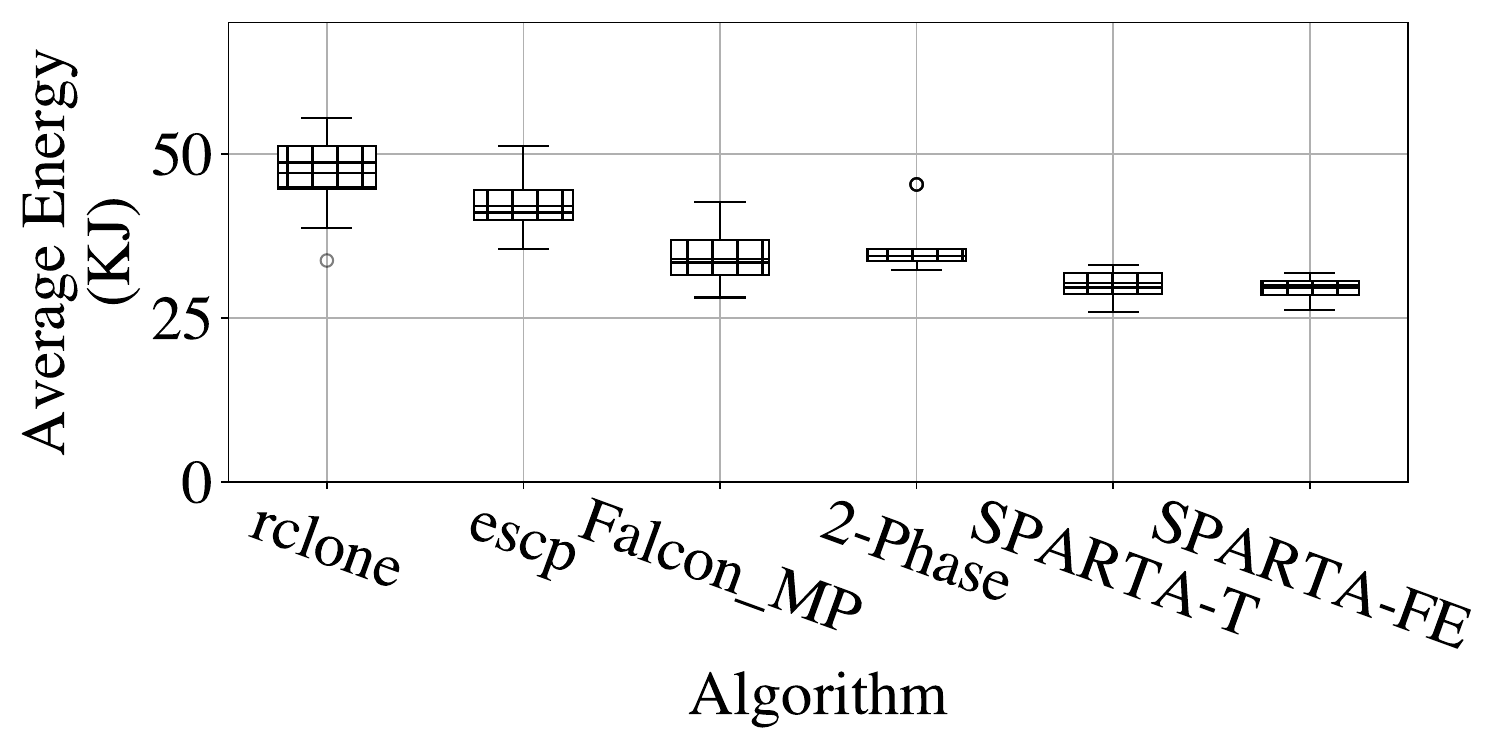}
    \caption{\small Energy in Cloudlab}
    \label{fig:cloudlab_en}
  \end{subfigure}
  \begin{subfigure}[b]{0.32\textwidth}
    \includegraphics[width=\textwidth]{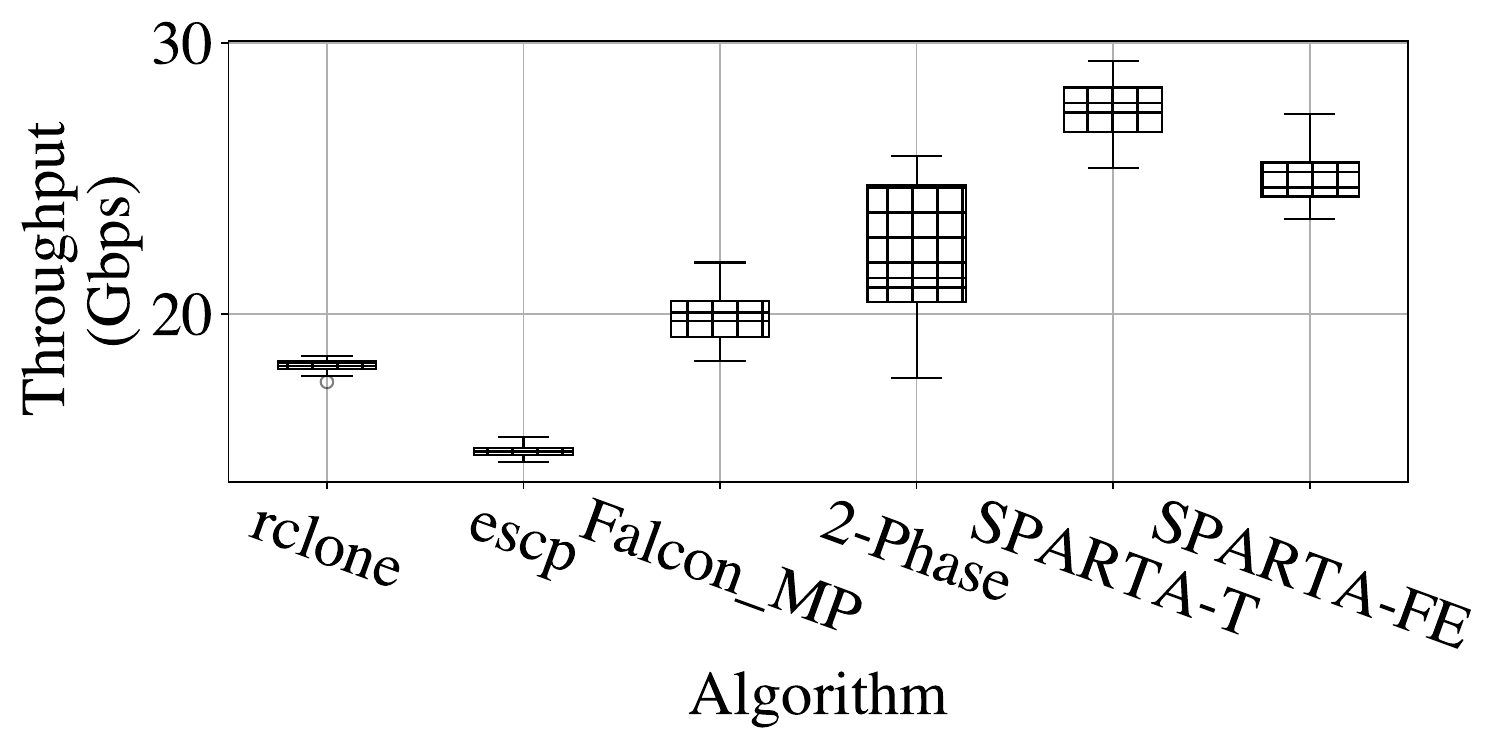}
    \caption{\small Throughput in Fabric}
    \label{fig:fabric_th}
  \end{subfigure}
  \caption{\small
Throughput and energy usage in Chameleon, CloudLab, and FABRIC (1\,TB of 1\,GB files) using six methods: \texttt{rclone}, \texttt{escp}, \texttt{Falcon\_MP}, \texttt{2-phase}, \texttt{SPARTA-T} and \texttt{SPARTA-FE}. The static tools (\texttt{rclone}, \texttt{escp}) fix $(cc,p)=(4,4)$, while \texttt{Falcon\_MP} updates $(cc,p)$ via gradient descent from a baseline $(cc,p)=(1,1)$. \texttt{SPARTA-T} and \texttt{SPARTA-FE} build on reinforcement learning . \texttt{2-phase} typically uses historical logs, but here starts from midpoint settings in the absence of extensive data. Due to a lack of hardware counters in FABRIC, only throughput is reported for that testbed. 
}
  \label{fig:comparisonAlgos}
\end{figure*}

We compare two static tools (\texttt{rclone} and \texttt{escp}), an online optimizer (\texttt{Falcon\_MP}~\cite{falcon-2023}), a historical-data-driven method called \texttt{2-phase}~\cite{twoPhase}, and our two DRL-based approaches (\texttt{SPARTA-T} and \texttt{SPARTA-FE}). The \texttt{2-phase} model typically mines prior historical data to guide parameter tuning in real time, but because we did not have historical datasets in our testbed setup, we initialized it from a midpoint range of concurrency ($cc$) and parallelism ($p$). For both DRL \texttt{SPARTA} agents, we initially set $cc$ and $p$ at a midpoint within the concurrency/parallelism range. If prior knowledge about promising throughput at certain settings is available, these initial values can instead be chosen accordingly. Meanwhile, \texttt{Falcon\_MP} starts from a baseline configuration and uses gradient descent to tune concurrency and parallelism, while \texttt{rclone} and \texttt{escp} maintain the same parameters for the duration of each session. \\

\vspace{-10mm}
\subsection{Experimental Setup}

We conduct experiments on three distinct cloud environments with different network connectivity to illustrate the effectiveness of our approach:

\textbf{Chameleon Cloud:} 
    We deploy the sender and receiver on \textit{gpu\_p100} nodes at TACC, featuring Intel Xeon E5-2670 v3 processors (2 CPUs, 48 threads total), 128\,GB RAM, and two NVIDIA P100 GPUs. Each node also includes 400--1000\,GB of local storage and a 10\,Gbps NIC (Broadcom NetXtreme II BCM57800). The WAN path between TACC and the University of Chicago sites is shared and can support up to 10\,Gbps of throughput.

\textbf{CloudLab:} Our experiments here utilized two node types:
        \textit{c6525-100g} (Utah Site): AMD EPYC 7402P with 48 cores @2.8\,GHz, 128\,GB RAM, 3200\,GB local disk, and a 25\,Gbps NIC (bandwidth externally capped at 25\,Gbps).
        \textit{d7525} (Wisconsin Site): AMD EPYC 64 cores @3.0\,GHz, 128\,GB RAM, 2560\,GB local disk, and a 200\,Gbps NIC (again limited to 25\,Gbps for the WAN).
    We place the sender and receiver at Utah and Wisconsin, respectively, to create a wide-area transfer scenario.

    \textbf{FABRIC:} 
    We use virtual machine (VM) instances at the Princeton and Utah sites, each equipped with 32 CPU cores, 128\,GB of RAM, and 2560\,GB of disk. Each VM accesses a \textit{ConnectX\_6} 100\,Gbps NIC, though the effective WAN bandwidth was near 30\,Gbps. Because these are virtualized environments, no direct hardware counters are available for energy measurements. Consequently, we only report throughput for FABRIC experiments.

All transfers are performed using Apache servers on the sender side and memory-to-memory transfers at the destination, with TCP CUBIC as the transport protocol across all sites. In Chameleon and CloudLab, we measure energy consumption using Intel’s Running Average Power Limit (RAPL)~\cite{jamil-ICCCN-2022}, subtracting each system’s baseline power to isolate the energy used for data transfers. Since FABRIC lacks direct hardware counters for energy measurements, we only report throughput.
\vspace{-5mm}
\subsection{Performance Across Testbeds}
Figure~\ref{fig:comparisonAlgos} summarizes throughput and energy usage at the Chameleon Cloud and CloudLab, and throughput only at the FABRIC testbeds. 


\texttt{rclone} and \texttt{escp} rely on static $(cc, p) = (4,4)$ configuration, averaging around 4--6\,Gbps transfer throughput. \texttt{Falcon\_MP}, which starts from a baseline and uses gradient descent, reaches about 8\,Gbps after multiple iterations. \texttt{SPARTA-T} and \texttt{SPARTA-FE} adapt their parameters more flexibly, often hitting 9--10\,Gbps. Meanwhile, \texttt{2-phase}, which depends on extensive historical logs to guide parameter choices, could not fully exploit its offline modeling here and thus settles near 7\,Gbps. 

\texttt{rclone} and \texttt{escp} exhibit relatively high energy usage due to underutilized link capacity resulting in a prolonged transfer time. \texttt{Falcon\_MP} does better than \texttt{rclone} and \texttt{escp} due to improved throughput and smaller transfer time.  \texttt{SPARTA-T} and \texttt{SPARTA-FE} maintains the lowest energy expenditure by directly optimizing for throughput-to-energy ratios. \texttt{2-phase} remains in a mid-range energy profile, reflecting its partial adaptation without an extensive historical dataset. 

\subsubsection{CloudLab (Figures~\ref{fig:cloudlab_th} and~\ref{fig:cloudlab_en})}
Constrained by a 25\,Gbps link, \texttt{SPARTA-T} and \texttt{SPARTA-FE} each manage 22--24\,Gbps transfer throughput on average, surpassing \texttt{rclone} and \texttt{escp} (16--18\,Gbps). \texttt{Falcon\_MP} again requires multiple steps to approach 20\,Gbps. Lacking its usual historical data, \texttt{2-phase} settles near 14\,Gbps, trailing both the DRL based \texttt{SPARTA} agents. 

\texttt{SPARTA-FE} continues to post the lowest energy usage, validating its energy-aware reward function. \texttt{SPARTA-T} uses slightly more energy compared to  \texttt{SPARTA-FE}. \texttt{2-phase} shows moderate energy consumption in the absence of extensive logs, while \texttt{Falcon\_MP}, starting from a baseline, experiences longer convergence and resulting higher energy usage.

\subsubsection{FABRIC (Figure~\ref{fig:fabric_th})}
No hardware counters for energy measurement are available in FABRIC, so we only report the transfer throughput. Despite a nominal 100\,Gbps link and a 56\,ms RTT, practical factors such as shared NIC among VMs limit attainable average throughput to roughly 28\,Gbps (Figure~\ref{fig:fabric_th}). Within these bounds, both DRL \texttt{SPARTA} methods achieve 20--25\,Gbps, while \texttt{Falcon\_MP} converges more slowly. \texttt{2-phase} also lags behind the DRL based \texttt{SPARTA} agents due to lack of historical transfer data.

Overall, the DRL-based \texttt{SPARTA} agents outperform static, online optimizers and methods dependent on transition logs (\texttt{rclone}, \texttt{escp}, \texttt{Falcon\_MP}, \texttt{2-phase}) by continuously adjusting concurrency ($cc$) and parallelism ($p$) in near real time. In particular, \texttt{SPARTA-FE} achieves a lower overall energy footprint but delivers slightly less throughput compared to \texttt{SPARTA-T}. This difference arises because \texttt{SPARTA-FE} includes packet loss rate directly in its reward signal, making it more conservative; it reacts quickly to congestion by reducing parameters to avoid excessive packet loss and ensure fair bandwidth allocation. In contrast, \texttt{SPARTA-T} relies on the throughput-to-energy ratio, which typically drops only after throughput has already decreased (often following packet loss), so it has more time to maintain higher throughput—even though this leads to increased energy usage. Meanwhile, \texttt{Falcon\_MP} needs multiple gradient-descent steps from its baseline to converge, and \texttt{2-phase}, which typically exploits extensive historical logs, cannot fully leverage its modeling without such data—leading both to lag behind the DRL-based \texttt{SPARTA} agents. These results underscore the advantages of learning-based, multi-parameter adaptation in high-bandwidth, dynamic network environments. 

\begin{figure}[t]
    \begin{subfigure}[b]{0.48\textwidth}
    \includegraphics[width=\textwidth]{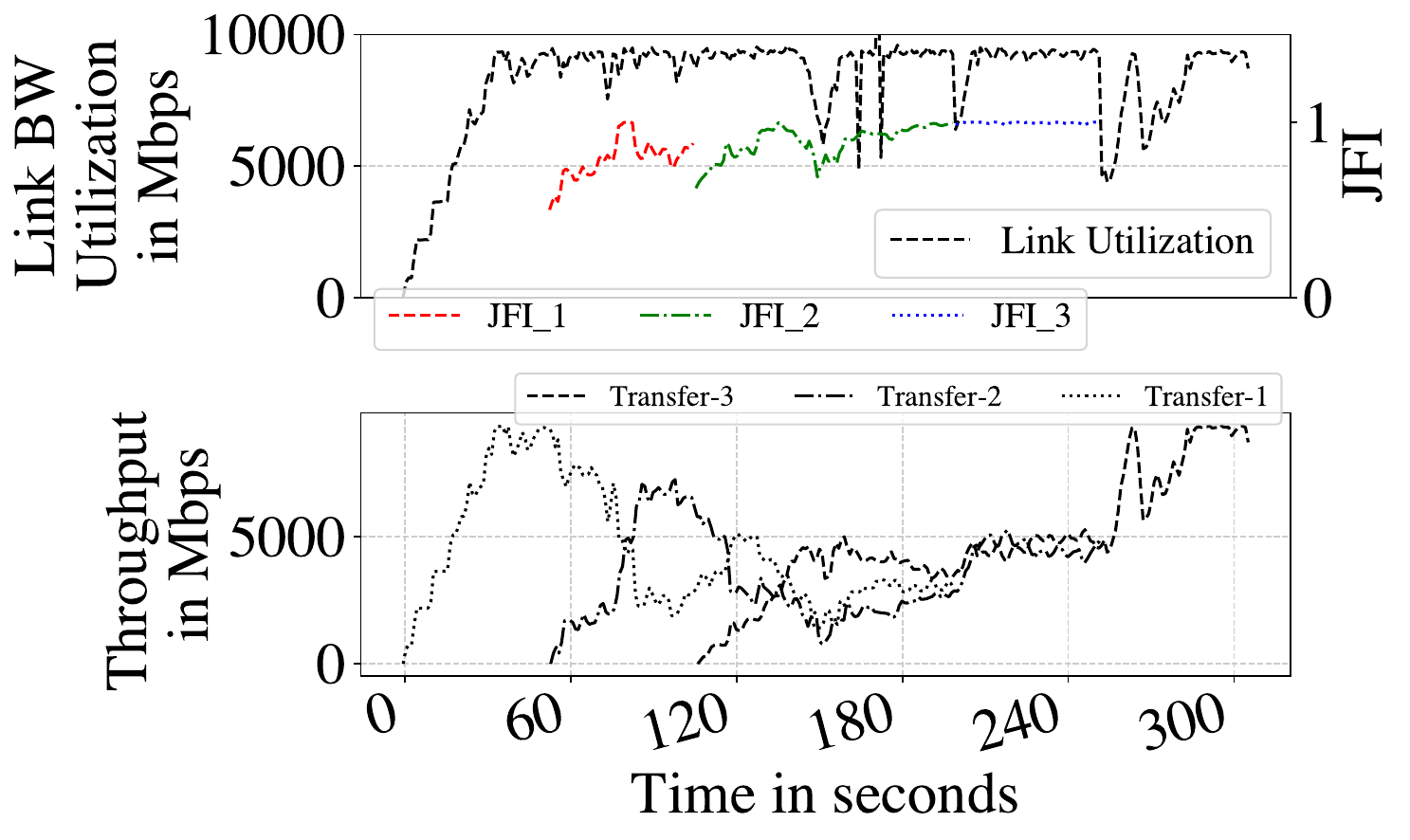}
    \caption{Transfers sharing a 10G link each running SPARTA-T}
    \label{fig:fairness-DRL-EE}
  \end{subfigure}    
\begin{subfigure}[b]{0.48\textwidth}
    \includegraphics[width=\textwidth]{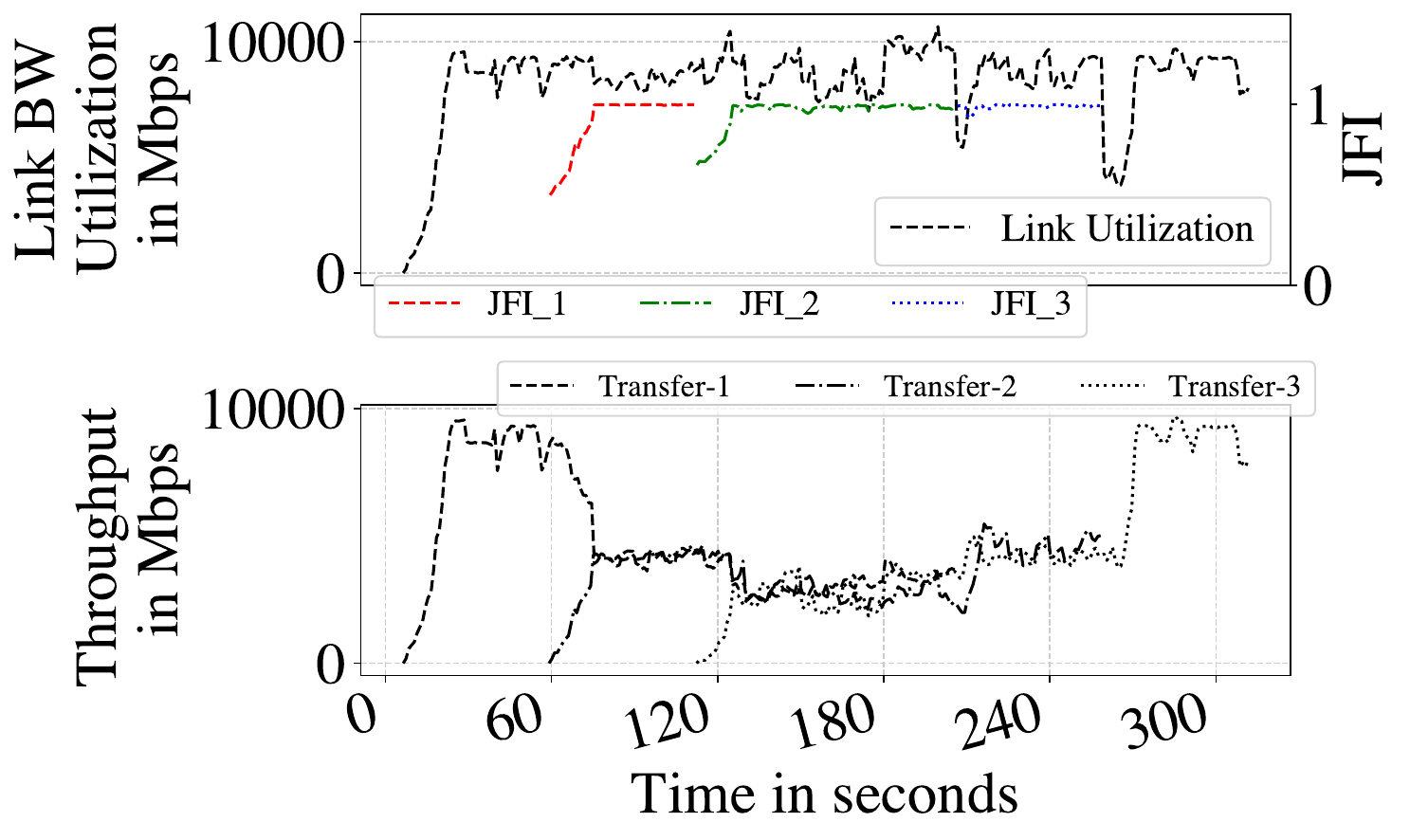}
    \caption{Transfers sharing a 10G link each running SPARTA-FE}
    \label{fig:fairness-DRL-F}
  \end{subfigure}
    \begin{subfigure}[b]{0.45\textwidth}
        \hspace{4mm}
        \includegraphics[width=\textwidth]{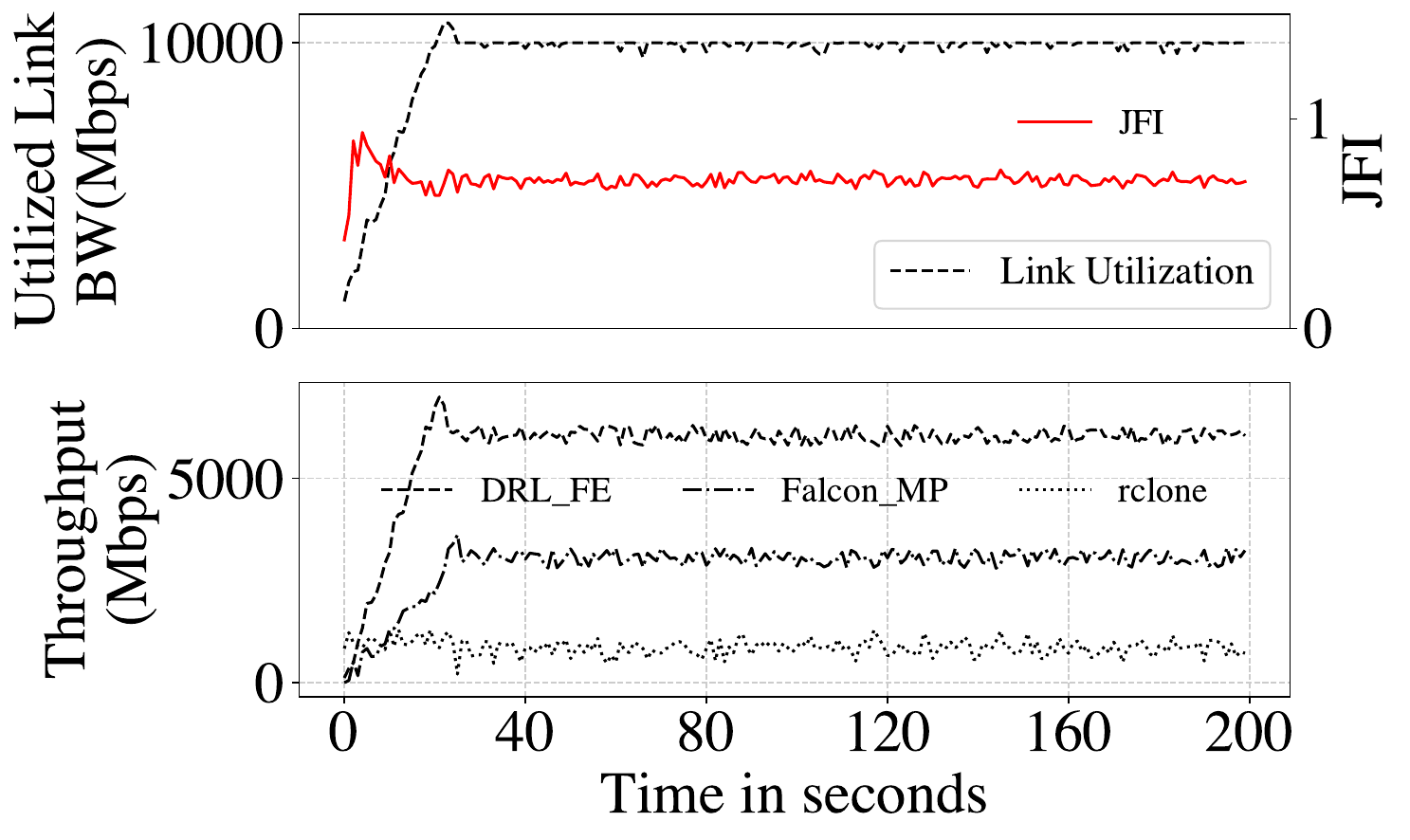}
        \caption{Transfers sharing a 10G link each running a different algorithm}
        \label{fig:fairness-DRL-rclone-falcon}
      \end{subfigure}
      \caption{\small Comparison of transfer performance in Chameleon cloud between TACC and UC sites. (a) Shows three transfers running separate \texttt{SPARTA-T} for a 10G shared link. (b) Shows three transfers running separate \texttt{SPARTA-FE} with the same shared environment. \texttt{SPARTA-FE} achieves better fairness compared to \texttt{SPARTA-T} because of the inclusion of packet loss rate in the reward function, which makes \texttt{SPARTA-FE} share the available resources more fairly. 
      }
    
  \label{fig:comparisionAlgos-fairness}
\end{figure}
\vspace{-5mm}
\subsection{Fairness in Concurrent Transfers}
\label{sec:fairness}

In many real-world deployments, multiple data flows share the same bottleneck link, making fairness essential for efficient resource allocation. To assess this, we launch concurrent transfers using different optimizers. Figures~\ref{fig:fairness-DRL-EE}, \ref{fig:fairness-DRL-F}, and \ref{fig:fairness-DRL-rclone-falcon} show three representative scenarios:
\begin{itemize}
    \item (a) Three transfers running \texttt{SPARTA-T} \emph{Throughput-Focused Energy Efficiency} objective (T/E Reward, see Eq.~\ref{eq:reward_def} ).
    \item (b) Three transfers running \texttt{SPARTA-FE} \emph{Fairness and Efficiency} objective (F\&E Reward, see Eq.~\ref{eq:reward_def} ).
    \item (c) A mixed scenario with one transfer each using \texttt{SPARTA-FE}, \texttt{Falcon\_MP}, and \texttt{rclone}.
\end{itemize}


\subsubsection{Jain’s Fairness Index (JFI)}
We quantify fairness using Jain’s Fairness Index (JFI)~\cite{jain1998quantitative}, which measures how evenly bandwidth is distributed among $n$ concurrent flows with throughputs $\{T_1, T_2, \dots, T_n\}$:
3w3444444\begin{equation}
    \text{JFI} = \frac{(\sum_{k=1}^{n} T_k)^2}{n \sum_{k=1}^{n} (T_k)^2}.
\end{equation}
A JFI close to 1 indicates nearly perfect fairness; lower values signify larger throughput imbalances among flows.

\subsubsection{\texttt{SPARTA-T} (T/E Reward) Fairness (Figure~\ref{fig:fairness-DRL-EE})}
We begin by running three concurrent transfers using \texttt{SPARTA-T}, which implements the \emph{Throughput-Focused Energy Efficiency} (T/E) Reward. As each flow seeks to maximize its throughput-to-energy ratio, they converge to different throughput levels depending on arrival times. Consequently, the overall Jain’s Fairness Index (JFI) remains moderate because the T/E Reward does not heavily penalize bandwidth imbalances. Moreover, \texttt{SPARTA-T} can show larger fluctuations in JFI when multiple flows share the network, since it adjusts concurrency and parallelism primarily to optimize throughput per energy. Once a flow completes or detects diminishing energy returns, other flows opportunistically claim the freed capacity, yielding strong per-bit energy savings but potentially fluctuate fairness score among simultaneous transfers.


\subsubsection{\texttt{SPARTA-FE} (F\&E Reward) Fairness (Figure~\ref{fig:fairness-DRL-F})}
Next, we repeat the three-transfer experiment with \texttt{SPARTA-FE}, which employs the \emph{Fairness and Efficiency} (F\&E) Reward. This reward explicitly factors in packet loss to enforce fairness while also indirectly reducing wasted energy from congestion. After a brief exploration phase, all flows converge to an equal share of the link, yielding a significantly higher JFI. When any flow ends, the remaining flows dynamically scale their parameters to use the newly available bandwidth, while still preventing congestion. Because packet loss directly impacts the F\&E Reward, \texttt{SPARTA-FE} responds faster to incipient congestion than T/E-based approaches, resulting in more stable fairness when multiple flows are active.

\subsubsection{Mixed Algorithms Fairness (Figure~\ref{fig:fairness-DRL-rclone-falcon})}
Finally, we test a scenario with three concurrent data transfers, each using a different approach:
\begin{itemize}
    \item \texttt{SPARTA-FE} (F\&E Reward),
    \item \texttt{Falcon\_MP} (online concurrency/parallelism tuning via gradient descent),
    \item \texttt{rclone} (static concurrency=4, parallelism=4).
\end{itemize}
In Figure~\ref{fig:fairness-DRL-rclone-falcon}, \texttt{SPARTA-FE} and \texttt{Falcon\_MP} both start from similar parameter settings. However, \texttt{SPARTA-FE} quickly and intelligently adjusts $(cc,p)$, achieving high throughput sooner. As \texttt{Falcon\_MP} gradually ramps up its concurrency, \texttt{SPARTA-FE} reduces its own parameters slightly to accommodate the new flows. By contrast, \texttt{rclone} remains at its baseline setting. Overall, the JFI remains high, indicating that \texttt{SPARTA-FE} maintains a balanced share of the link after convergence.

In summary, \texttt{SPARTA-T} optimizes throughput per unit energy, sometimes allowing throughput imbalances to persist. Conversely, \texttt{SPARTA-FE} factors packet loss directly into its reward, actively discouraging congestion and achieving higher fairness across flows. Even in the mixed scenario, \texttt{SPARTA-FE} adapts its parameters to accommodate incoming flows or reclaim unused capacity, resulting in consistently higher JFI than T/E-based or static approaches.

\section{Conclusion}
\label{sec:conclusion}

In this paper, we present \texttt{SPARTA}, a multi-parameter deep reinforcement learning (DRL) framework that automatically tunes concurrency (\(cc\)) and parallelism (\(p\)) to boost data transfer performance in high-speed networks while keeping the energy consumption low. Using real-world state transition data, we design two agent variants: \texttt{SPARTA-T} with a \emph{throughput-focused energy efficiency (T/E) reward} and \texttt{SPARTA-FE} with a \emph{fairness and efficiency (F\&E) Reward}. These reward signals enable the agents to balance throughput with energy usage while ensuring fair bandwidth sharing. Tested on multiple testbeds, \texttt{SPARTA} consistently outperforms both baseline and state-of-the-art methods, requiring only minimal configuration and showcasing the advantages of adaptive resource allocation.

\texttt{SPARTA} trains DRL agents using energy- and fairness-aware reward signals, allowing them to pause and resume transfer threads based on real-time network conditions. This flexibility not only speeds up data transfers but also prevents overloading resources and lowers energy use. To speed up training, we introduce an emulation environment that stores transition logs from initial real-world training episodes. By learning from these logs, the agents quickly discover optimal actions without the high costs of lengthy real-world transfers. After training, they intelligently adjust $cc$ and $p$ to avoid resource congestion during busy periods and to fully use available bandwidth during quieter times, all without sacrificing fairness.

Our results show that DRL-based resource allocation can build more sustainable, high-performing data transfer infrastructures that save energy, increase throughput, and maintain fairness.
Although our experiments focus on multiple concurrent transfers on shared links, future work could scale this approach to thousands of transfers across geographically distributed environments. This might involve multi-agent DRL or centralized optimization strategies and also incorporate other transport protocols, such as TCP BBR ~\cite{BBR2016}.

\section*{Acknowledgments}
This project is in part sponsored by the National Science Foundation (NSF) under award number OAC-2313061.


\bibliographystyle{IEEEtran}
\bibliography{references}

@inproceedings{nine2018greendataflow,
  title={Greendataflow: Minimizing the energy footprint of global data movement},
  author={Nine, MD SQ Zulkar and Di Tacchio, Luigi and Imran, Asif and Kosar, Tevfik and Bulut, M Fatih and Hwang, Jinho},
  booktitle={2018 IEEE International Conference on Big Data (Big Data)},
  pages={335--342},
  year={2018},
  organization={IEEE}
}

@article{shalf2020future,
  title={The future of computing beyond Moore’s Law},
  author={Shalf, John},
  journal={Philosophical Transactions of the Royal Society A},
  volume={378},
  number={2166},
  pages={20190061},
  year={2020},
  publisher={The Royal Society Publishing}
}

@MISC{Cisco2023,
  author        = {},
  title         = "{Global IP Traffic Forecast and Methodology, 2018--2023 (White Paper)}",
  howpublished  = {Cisco Systems},
  key           = {},
  month         = {},
  year          = {2023},
  note          = {}
}

@article{aujoux2021estimating,
  title={Estimating the carbon footprint of the GRAND project, a multi-decade astrophysics experiment},
  author={Aujoux, Clarisse and Kotera, Kumiko and Blanchard, Odile},
  journal={Astroparticle Physics},
  volume={131},
  pages={102587},
  year={2021},
  publisher={Elsevier}
}

@misc{lorincz2019greener,
  title={Greener, energy-efficient and sustainable networks: State-of-the-art and new trends},
  author={Lorincz, Josip and Capone, Antonio and Wu, Jinsong},
  journal={Sensors},
  volume={19},
  number={22},
  pages={4864},
  year={2019},
  publisher={MDPI}
}

@ARTICLE{twoPhase,
  author={Nine, MD S Q Zulkar and Kosar, Tevfik},
  journal={IEEE Transactions on Parallel and Distributed Systems}, 
  title={A Two-Phase Dynamic Throughput Optimization Model for Big Data Transfers}, 
  year={2021},
  volume={32},
  number={2},
  pages={269-280},
}

@article{kim2015highly,
  title={A highly-accurate and low-overhead prediction model for transfer throughput optimization},
  author={Kim, JangYoung and Yildirim, Esma and Kosar, Tevfik},
  journal={Cluster Computing},
  volume={18},
  pages={41--59},
  year={2015},
  publisher={Springer}
}

@book{kosar2005data,
  title={Data placement in widely distributed systems},
  author={Kosar, Tevfik},
  year={2005},
  publisher={The University of Wisconsin-Madison}
}

@inproceedings{balman2007data,
  title={Data scheduling for large scale distributed applications},
  author={Balman, Mehmet and Kosar, Tevfik},
  booktitle={the 5th ICEIS Doctoral Consortium, ICEIS’07. Funchal, Madeira-Portugal},
  year={2007}
}

@INPROCEEDINGS{jamil-ICCCN-2022,
  author={Jamil, Hasibul and Rodolph, Lavone and Goldverg, Jacob and Kosar, Tevfik},
  booktitle={ICCCN}, 
  title={Energy-Efficient Data Transfer Optimization via Decision-Tree Based Uncertainty Reduction}, 
  year={2022},
  volume={},
  number={},
  pages={1-10},
}

@INPROCEEDINGS{Dong-2005,
  author={Dong Lu and Yi Qiao and Dinda, P.A. and Bustamante, F.E.},
  booktitle={IPDPS}, 
  title={Modeling and taming parallel TCP on the wide area network}, 
  year={2005},
  volume={},
  number={},
  doi={10.1109/IPDPS.2005.291}}

@article{Yildirim-TCC2016,
author = {Yildirim, Esma and Arslan, Engin and Kim, Jangyoung and Kosar, Tevfik},
title = {Application-Level Optimization of Big Data Transfers through Pipelining, Parallelism and Concurrency},
year = {2016},
issue_date = {January 2016},
publisher = {IEEE Computer Society Press},
address = {Washington, DC, USA},
volume = {4},
number = {1},
issn = {2168-7161},
journal = {IEEE Trans. Cloud Comput.},
month = jan,
pages = {63–75},
numpages = {13}
}

@article{belkhir2018assessing,
  title={Assessing ICT global emissions footprint: Trends to 2040 \& recommendations},
  author={Belkhir, Lotfi and Elmeligi, Ahmed},
  journal={Journal of cleaner production},
  volume={177},
  pages={448--463},
  year={2018},
  publisher={Elsevier}
}

@article{andrae2015global,
  title={On global electricity usage of communication technology: trends to 2030},
  author={Andrae, Anders SG and Edler, Tomas},
  journal={Challenges},
  volume={6},
  number={1},
  pages={117--157},
  year={2015},
  publisher={Multidisciplinary Digital Publishing Institute}
}

@INPROCEEDINGS{hacker2002,
  author={Hacker, T.J. and Athey, B.D. and Noble, B.},
  booktitle={IPDPS}, 
  title={The end-to-end performance effects of parallel TCP sockets on a lossy wide-area network}, 
  year={2002},
  volume={},
  number={},
}

@INPROCEEDINGS{cubic2008,
author = {Ha, Sangtae and Rhee, Injong and Xu, Lisong},
title = {CUBIC: A New TCP-Friendly High-Speed TCP Variant},
year = {2008},
booktitle = {SIGOPS Oper. Syst. Rev.},
issue_date = {2008},
volume = {42},
number = {5},
pages = {64–74},
numpages = {11}
}

@INPROCEEDINGS{mathis1997,
author = {Mathis, Matthew and Semke, Jeffrey and Mahdavi, Jamshid and Ott, Teunis},
title = {The Macroscopic Behavior of the TCP Congestion Avoidance Algorithm},
year = {1997},
issue_date = {July 1997},
volume = {27},
number = {3},
issn = {0146-4833},
booktitle = {SIGCOMM Comput. Commun. Rev.},
month = {jul},
pages = {67–82},
numpages = {16}
}

@inproceedings{keahey2020lessons,
  title={Lessons Learned from the Chameleon Testbed},
  author={Kate Keahey and Jason Anderson and Zhuo Zhen and et al.},
  booktitle={Proceedings of USENIX ATC'20},
    month={July},
  year={2020}
}

@INPROCEEDINGS{di2019cross,
  author={Di Tacchio, Luigi and Nine, MD S Q Zulkar and Kosar, Tevfik and Bulut, Muhammed Fatih and Hwang, Jinho},
  booktitle={2019 IEEE CLOUD}, 
  title={Cross-Layer Optimization of Big Data Transfer Throughput and Energy Consumption}, 
  year={2019},
 }

@INPROCEEDINGS{Rodolph2021,
  author={Rodolph, Lavone and Zulkar Nine, MD S Q and Di Tacchio, Luigi and Kosar, Tevfik},
  booktitle={IEEE ICC 2021}, 
  title={Energy-saving Cross-layer Optimization of Big Data Transfer Based on Historical Log Analysis}, 
  year={2021},
  volume={},
  number={},
  pages={1-7},
 }

@article{BBR2016,
author = {Cardwell, Neal and Cheng, Yuchung and Gunn, C. Stephen and Yeganeh, Soheil Hassas and Jacobson, Van},
title = {BBR: Congestion-Based Congestion Control: Measuring Bottleneck Bandwidth and Round-Trip Propagation Time},
year = {2016},
issue_date = {September-October 2016},
publisher = {Association for Computing Machinery},
address = {New York, NY, USA},
issn = {1542-7730},
doi = {10.1145/3012426.3022184},
journal = {Queue},
month = {oct},
}

@inproceedings {PCC-Vivace,
author = {Mo Dong and Tong Meng and Doron Zarchy and Engin Arslan and Yossi Gilad and Brighten Godfrey and Michael Schapira},
title = {{PCC} Vivace: {Online-Learning} Congestion Control},
booktitle = {NSDI 18},
year = {2018},
isbn = {978-1-939133-01-4},
address = {Renton, WA},
}

@ARTICLE{qtcp,
  author={Li, Wei and Zhou, Fan and Chowdhury, Kaushik Roy and Meleis, Waleed},
  journal={IEEE Trans. on Network Science and Engineering}, 
  title={QTCP: Adaptive Congestion Control with Reinforcement Learning}, 
  year={2019},
  volume={6},
  number={3},
  pages={445-458},
  }

@article{DataCenterCongestionControl-2022,
  author    = {Chen Tessler and
               Yuval Shpigelman and
               Gal Dalal and
               Amit Mandelbaum and
               Doron Haritan Kazakov and
               Benjamin Fuhrer and
               Gal Chechik and
               Shie Mannor},
  title     = {Reinforcement Learning for Datacenter Congestion Control},
  journal   = {CoRR},
  volume    = {abs/2102.09337},
  year      = {2021},
  eprinttype = {arXiv},
  eprint    = {2102.09337},
  bibsource = {dblp computer science bibliography, https://dblp.org}
}

@InProceedings{nathan-jay2019,
  title = 	 {A Deep Reinforcement Learning Perspective on Internet Congestion Control},
  author =       {Jay, Nathan and Rotman, Noga and Godfrey, Brighten and Schapira, Michael and Tamar, Aviv},
  booktitle = 	 {Proceedings of ICML},
  pages = 	 {3050--3059},
  year = 	 {2019},
  volume = 	 {97},
  month = 	 {09--15 Jun},
 }

@misc{jain1998quantitative,
      title={A Quantitative Measure Of Fairness And Discrimination For Resource Allocation In Shared Computer Systems}, 
      author={R. Jain and D. Chiu and W. Hawe},
      year={1998},
      eprint={cs/9809099},
      archivePrefix={arXiv},
      primaryClass={cs.NI}
}

@inproceedings{Brooks:2000:WFA:339647.339657,
 author = {Brooks, David and Tiwari, Vivek and Martonosi, Margaret},
 title = {Wattch: a framework for architectural-level power analysis and optimizations},
 booktitle = {Proceedings of ISCA'00},
 year = {2000},
 location = {Vancouver, BC, Canada},
 pages = {83--94},
 numpages = {12},
 }

@article{rawson2004mempower,
  title={MEMPOWER: A simple memory power analysis tool set},
  author={Rawson, Freeman and Austin, IBM},
  journal={IBM Austin Research Laboratory},
  year={2004}
}

@inproceedings{zedlewski2003modeling,
  title={Modeling Hard-Disk Power Consumption.},
  author={Zedlewski, John and Sobti, Sumeet and Garg, Nitin and Zheng, Fengzhou and Krishnamurthy, Arvind and Wang, Randolph Y},
  booktitle={FAST 2003},
}

@inproceedings{gurumurthi2002using,
  title={Using complete machine simulation for software power estimation: The softwatt approach},
  author={Gurumurthi, Sudhanva and Sivasubramaniam, Anand and Irwin, Mary Jane and Vijaykrishnan, Narayanan and Kandemir, Mahmut},
  booktitle={Prpc. of 8th High-Performance Computer Architecture Symp.},
  pages={141--150},
  year={2002},
}

@inproceedings{contreras2005power,
  title={Power prediction for intel XScale{\textregistered} processors using performance monitoring unit events},
  author={Contreras, Gilberto and Martonosi, Margaret},
  booktitle={ ISLPED'05},
  pages={221--226},
  year={2005},
}

@inproceedings{economou2006full,
  title={Full-system power analysis and modeling for server environments},
  author={Economou, Dimitris and Rivoire, Suzanne and Kozyrakis, Christos and Ranganathan, Partha},
  booktitle={Proc. of Workshop on Modeling, Benchmarking, and Simulation},
  year={2006}
}

@article{fan2007power,
  title={Power provisioning for a warehouse-sized computer},
  author={Fan, Xiaobo and Weber, Wolf-Dietrich and Barroso, Luiz Andre},
  journal={ACM SIGARCH Computer Architecture News},
  volume={35},
  number={2},
  pages={13--23},
  year={2007},
  publisher={ACM}
}

@inproceedings{koller2010wattapp,
  title={WattApp: an application aware power meter for shared data centers},
  author={Koller, Ricardo and Verma, Akshat and Neogi, Anindya},
  booktitle={Proceedings of the 7th international conference on Autonomic computing},
  pages={31--40},
  year={2010},
  organization={ACM}
}

@article{rivoire2008comparison,
  title={A Comparison of High-Level Full-System Power Models.},
  author={Rivoire, Suzanne and Ranganathan, Parthasarathy and Kozyrakis, Christos},
  journal={HotPower},
  volume={8},
    year={2008}
}

@inproceedings{hasebe2010power,
  title={Power-saving in large-scale storage systems with data migration},
  author={Hasebe, Koji and Niwa, Tatsuya and Sugiki, Akiyoshi and Kato, Kazuhiko},
  booktitle={IEEE CloudCom 2010},
}

@article{vrbsky2013decreasing,
  title={Decreasing power consumption with energy efficient data aware strategies},
  author={Vrbsky, Susan V and Galloway, Michael and Carr, Robert and Nori, Rahul and Grubic, David},
  journal={FGCS},
  volume={29},
  number={5},
  pages={1152--1163},
  year={2013},
  publisher={Elsevier}
}

@inproceedings{Katz_2008, 
author = "G. Ananthanarayanan and R. Katz", 
title = "Greening the Switch", 
booktitle = "{In Proceedings of HotPower, December 2008}"  
}

@inproceedings{Mahadevan_2009, 
author = "P. Mahadevan and P. Sharma and S. Banerjee and P. Ranganathan", 
title = "A Power Benchmarking Framework for Network Devices", 
booktitle = "{In Proceedings of IFIP Networking, May 2009}"  
}

@inproceedings{Greenberg_2009, 
author = "A. Greenberg and J. Hamilton and D. Maltz and P. Patel", 
title = "The Cost of a Cloud: Research Problems in Data Center Networks", 
booktitle = "{In ACM SIGCOMM CCR, January 2009}"  
}

@inproceedings{alan2015energy,
  title={Energy-aware data transfer algorithms},
  author={Alan, Ismail and Arslan, Engin and Kosar, Tevfik},
  booktitle={Proceedings of SC'15},
  pages={1--12},
  year={2015}
}

@inproceedings{nine2023greennfv,
  title={GreenNFV: Energy-Efficient Network Function Virtualization with Service Level Agreement Constraints},
  author={Nine, Md SQ Zulkar and Kosar, Tevfik and Bulut, Muhammed Fatih and Hwang, Jinho},
  booktitle={Proceedings of SC'23},
  pages={1--12},
  year={2023}
}

@article{arslan2018big,
  title={Big data transfer optimization through adaptive parameter tuning},
  author={Arslan, Engin and Pehlivan, Bahadir A and Kosar, Tevfik},
  journal={Journal of Parallel and Distributed Computing},
  volume={120},
  pages={89--100},
  year={2018},
  publisher={Elsevier}
}

@article{arslan2018high,
  title={High-speed transfer optimization based on historical analysis and real-time tuning},
  author={Arslan, Engin and Guner, Kemal and Kosar, Tevfik},
  journal={IEEE Transactions on Parallel and Distributed Systems},
  volume={29},
  number={6},
  pages={1303--1316},
  year={2018},
  publisher={IEEE}
}

@inproceedings{guner2018energy,
  title={Energy-Efficient Mobile Network I/O},
  author={Guner, Kemal and Kosar, Tevfik},
  booktitle={Proceedings of IEEE GLOBECOM'18},
  pages={1--6},
  year={2018},
  organization={IEEE}
}

@INPROCEEDINGS{powerModelAlan,
  author={Alan, Ismail and Arslan, Engin and Kosar, Tevfik},
  booktitle={Proceedings of SC'15}, 
  title={Energy-aware data transfer algorithms}, 
  year={2015},
  volume={},
  number={},
  pages={1-12},
  }

@inproceedings {skyplane,
author = {Paras Jain and Sam Kumar and Sarah Wooders and Shishir G. Patil and Joseph E. Gonzalez and Ion Stoica},
title = {Skyplane: Optimizing Transfer Cost and Throughput Using {Cloud-Aware} Overlays},
booktitle = {Proc. of NSDI'23},
year = {2023},
isbn = {978-1-939133-33-5},
address = {Boston, MA},
pages = {1375--1389},
month = apr
}

@inproceedings{effingo,
  title={An exabyte a day: throughput-oriented, large scale, managed data transfers with Effingo},
  author={P{\'a}pay, Ladislav and Pustelnik, Jan and Rzadca, Krzysztof and Strack, Beata and Stradomski, Pawe{\l} and Wo{\l}owiec, Bart{\l}omiej and Zasadzinski, Michal},
  booktitle={Proceedings of ACM SIGCOMM'24},
  pages={970--982},
  year={2024}
}

@INPROCEEDINGS{blaze,
  author={Marru, Suresh and Freitag, Brian and Wannipurage, Dimuthu and Bommala, Uday Kumar and Pradier, Patrick and Demange, Christophe and Pantha, Nishan and Mukherjee, Tathagata and Rosich, Betlem and Monjoux, Eric and Ramachandran, Rahul},
  booktitle={Proceedings of IEEE CLOUD'23}, 
  title={Blaze: A High-Performance, Scalable, and Efficient Data Transfer Framework with Configurable and Extensible Features}, 
  year={2023},
  volume={},
  number={},
  pages={58-68},
  }

@ARTICLE{falcon-2023,
  author={Arifuzzaman, Md and Bockelman, Brian and Basney, James and Arslan, Engin},
  journal={IEEE Transactions on Parallel and Distributed Systems}, 
  title={Falcon: Fair and Efficient Online File Transfer Optimization}, 
  year={2023},
  volume={34},
  number={8},
  pages={2265-2278},
  }

@INPROCEEDINGS{jamil-2023-tcp,
  author={Jamil, Hasibul and Rodrigues, Elvis and Goldverg, Jacob and Kosar, Tevfik},
  booktitle={Proceedings of GLOBECOM'23},
  title={Learning to Maximize Network Bandwidth Utilization with Deep Reinforcement Learning}, 
  year={2023},
  volume={},
  number={},
  pages={3711-3716}
}

@article{globus,
author = {Allen, Bryce and Bresnahan, John and Childers, Lisa and Foster, Ian and Kandaswamy, Gopi and Kettimuthu, Raj and Kordas, Jack and Link, Mike and Martin, Stuart and Pickett, Karl and Tuecke, Steven},
title = {Software as a service for data scientists},
year = {2012},
issue_date = {February 2012},
publisher = {Association for Computing Machinery},
address = {New York, NY, USA},
volume = {55},
number = {2},
journal = {Commun. ACM},
month = feb,
pages = {81–88},
numpages = {8}
}

@ARTICLE{probData,
  author={Yun, Daqing and Wu, Chase Q. and Rao, Nageswara S. V. and Kettimuthu, Rajkumar},
  journal={IEEE/ACM Transactions on Networking}, 
  title={Advising Big Data Transfer Over Dedicated Connections Based on Profiling Optimization}, 
  year={2019},
  volume={27},
  number={6},
  pages={2280-2293},
  doi={10.1109/TNET.2019.2943884}}

@article{Mnih2015,
  author    = {V. Mnih and K. Kavukcuoglu and D. Silver and et al.},
  title     = {Human-level control through deep reinforcement learning},
  journal   = {Nature},
  year      = {2015},
  volume    = {518},
  pages     = {529--533},
  month     = {February},
  received   = {10 July 2014},
  accepted   = {16 January 2015},
  published  = {25 February 2015},
  issue_date = {26 February 2015}
}

@misc{schulman2017proximal,
  author    = {John Schulman and Filip Wolski and Prafulla Dhariwal and Alec Radford and Oleg Klimov},
  title     = {Proximal Policy Optimization Algorithms},
  howpublished = {arXiv:1707.06347},
  year      = {2017},
  url       = {https://arxiv.org/abs/1707.06347}
}

@misc{lillicrap2016continuous,
  author    = {Timothy P. Lillicrap and Jonathan J. Hunt and Alexander Pritzel and Nicolas Heess and Tom Erez and Yuval Tassa and David Silver and Daan Wierstra},
  title     = {Continuous control with deep reinforcement learning},
  howpublished = {arXiv:1509.02971},
  year      = {2016},
  url       = {https://arxiv.org/abs/1509.02971}
}

@inproceedings{kapturowski2019recurrent,
  author    = {Steven Kapturowski and Georg Ostrovski and John Quan and R{\'e}mi Munos and Will Dabney},
  title     = {Recurrent experience replay in distributed reinforcement learning},
  booktitle = {International Conference on Learning Representations (ICLR)},
  year      = {2019},
}

@inproceedings{hausknecht2015deep,
  author    = {Matthew J. Hausknecht and Peter Stone},
  title     = {Deep Recurrent Q-Learning for Partially Observable MDPs},
  booktitle = {2015 AAAI Fall Symposium Series},
  year      = {2015},
  }

@article{lloyd1982least,
  author    = {Stuart P. Lloyd},
  title     = {Least squares quantization in {PCM}},
  journal   = {IEEE Transactions on Information Theory},
  volume    = {28},
  number    = {2},
  pages     = {129--137},
  year      = {1982},
  publisher = {IEEE}
}

@misc{rpposb3,
  author       = {{Stable Baseline3}},
  title        = {{Stable Baseline3}},
  howpublished = {\url{https:$//sb3-contrib.readthedocs.io/en/master/modules/ppo_recurrent.html$}},
  note         = {Accessed: 2025-01-23}
}

\begin{IEEEbiography}[{\includegraphics[width=1in,height=1.25in,clip,keepaspectratio]{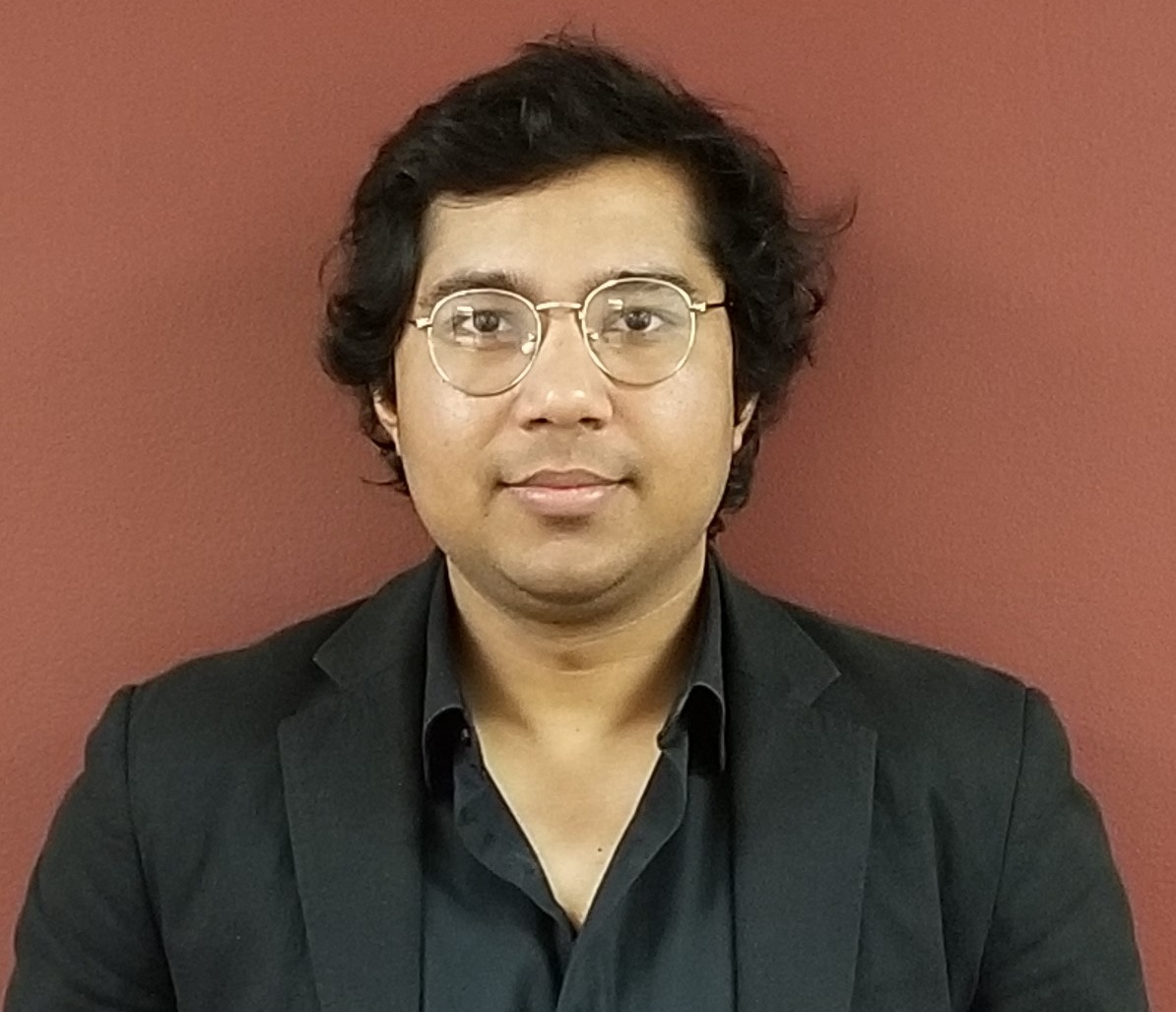}}]{Hasibul Jamil}
is a Ph.D. candidate in Computer Science and Engineering at the University at Buffalo, SUNY. He received his BS degree in Electrical and Electronic Engineering from Chittagong University of Engineering and Technology, Bangladesh, MS degree in Electrical and Computer Engineering from Southern Illinois University. He has interned at Pacific Northwest National Laboratory(PNNL), Argonne National Laboratory(ANL), and IBM TJ Watson Research Center. His research interests include High-performance systems, Distributed systems, sustainable computing, computer networks, and cloud systems. He is a IEEE student member. 
\end{IEEEbiography}

\vspace{-5mm}
\begin{IEEEbiography}[{\includegraphics[width=1in,height=1.25in,clip,keepaspectratio]{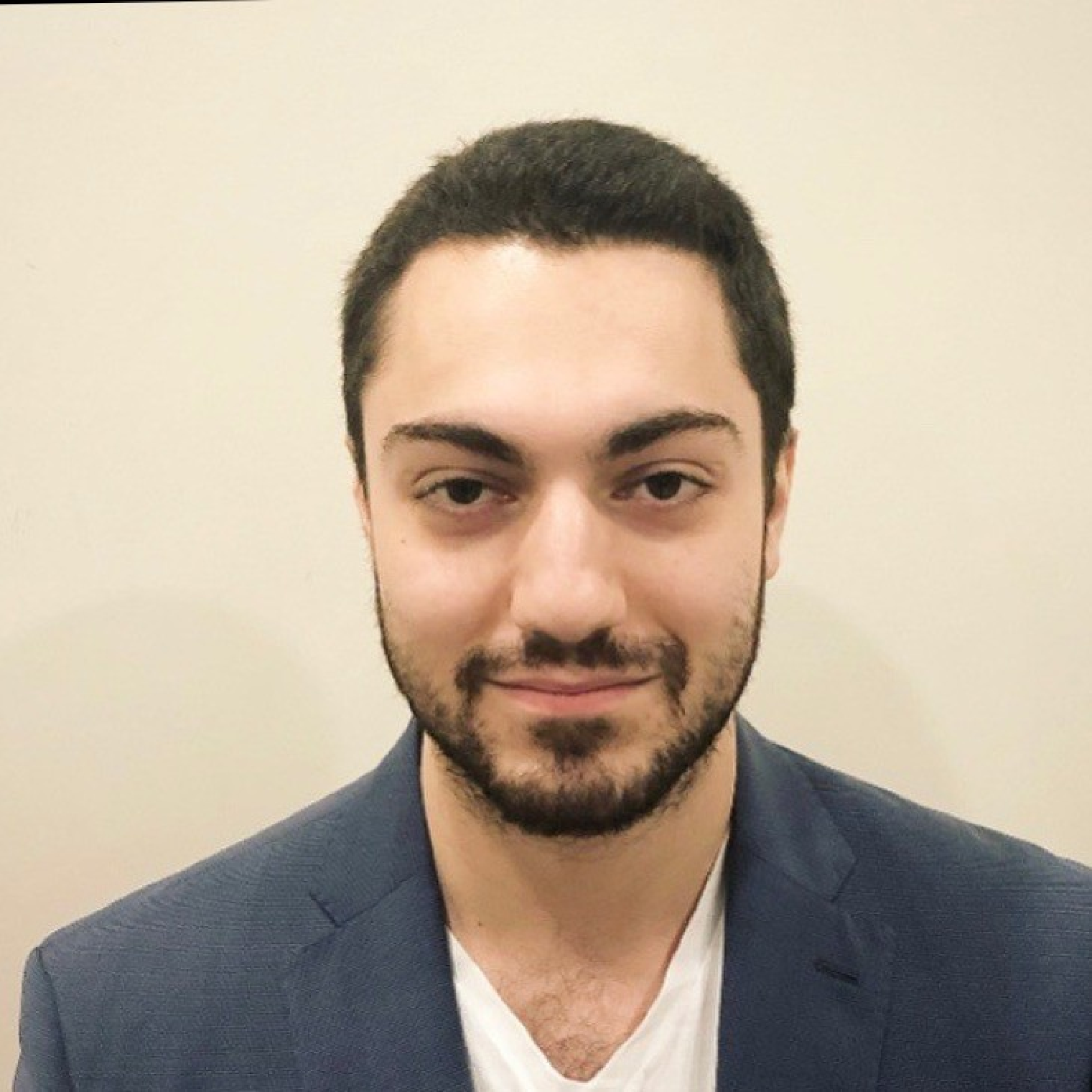}}]{Jacob Goldverg}
is a Ph.D. candidate in Computer Science and Engineering at the University at Buffalo, SUNY. He received his BS degree in Computer Science and Mathematics from the University at Buffalo, SUNY. He has interned at OverOps, University of Chicago, and Apple. His research interests include High-performance systems, sustainable computing, computer networks, and cloud systems.
\end{IEEEbiography}

\begin{IEEEbiography}[{\includegraphics[width=1in,height=1.25in]{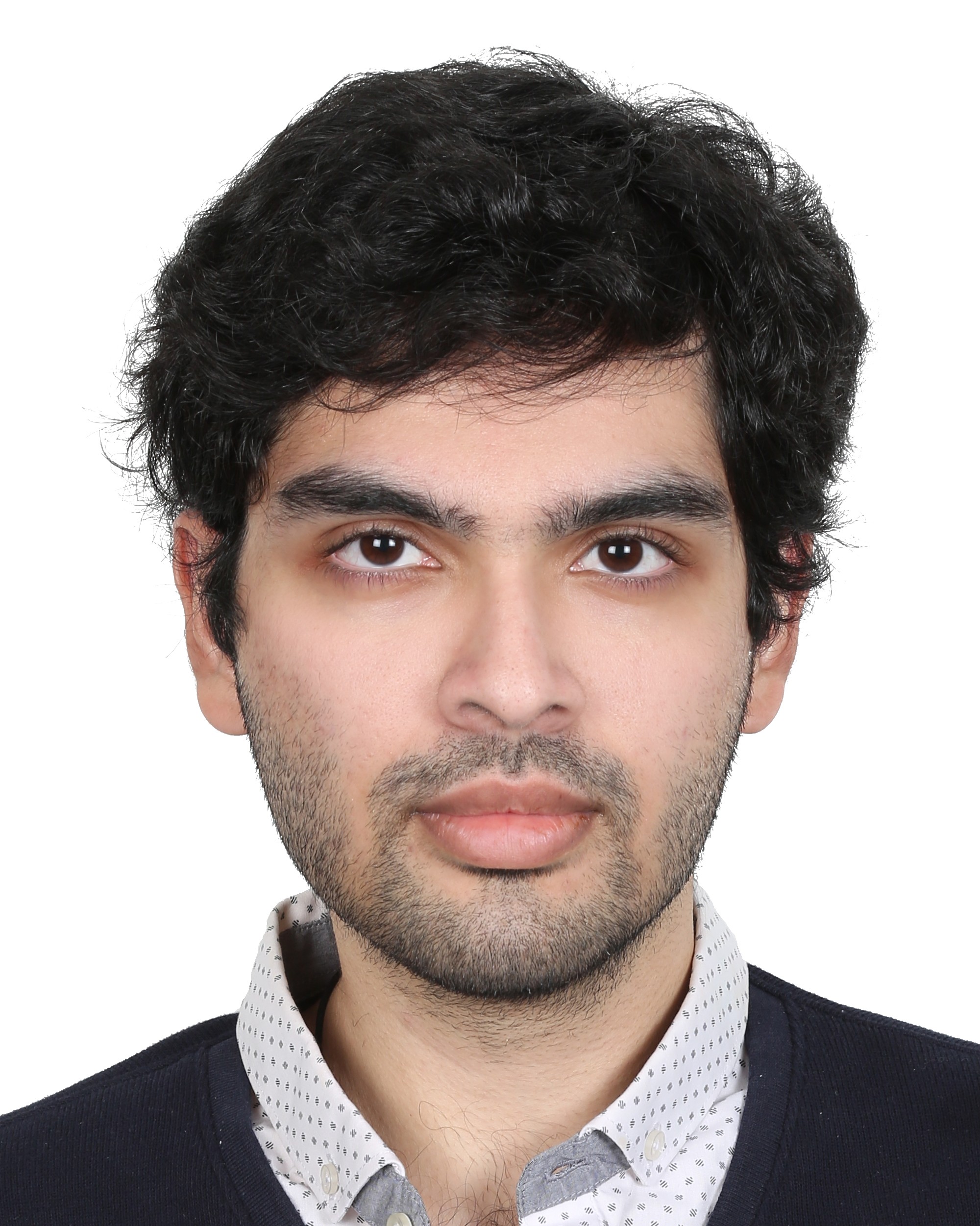}}]{Elvis Rodrigues} is a Ph.D. student in the Computer Science and Engineering Department at the University at Buffalo. He received his BS degree in Computer Science from the University of California, Los Angeles. His research interests include distributed systems, sustainable computing, scalability and performance of wide-area networked systems, and energy-efficient system optimization.
\end{IEEEbiography}

\begin{IEEEbiography}[{\includegraphics[width=1in,height=1.25in,clip,keepaspectratio]{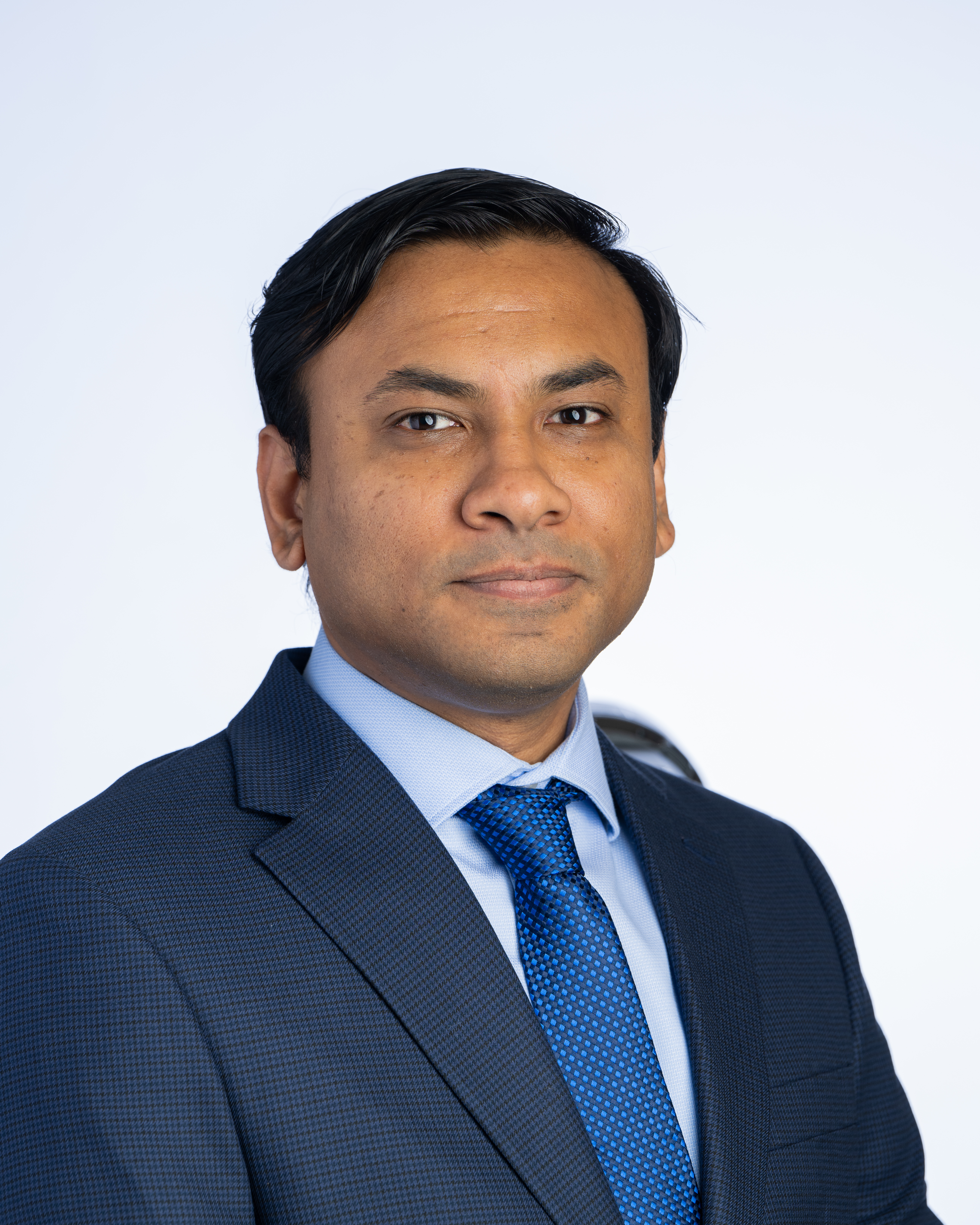}}]{MD S Q Zulkar Nine}
is an Assistant Professor of Computer Science at Tennessee Technological University. He received his BS degree in computer engineering from the Military Institute of Science and Technology, Dhaka, Bangladesh, his MS degree from North South University, Dhaka, Bangladesh, and his Ph.D. in Computer Science and Engineering from the University at Buffalo, SUNY. He also worked at IBM TJ Watson Research Center as a Research Intern. His research interest includes Distributed Machine Learning, High-performance networks, network and protocol optimization, Distributed systems, computer networks, and cloud systems. He is a member of the IEEE.
\end{IEEEbiography}

\begin{IEEEbiography}[{\includegraphics[width=1in,height=1.25in,clip,keepaspectratio]{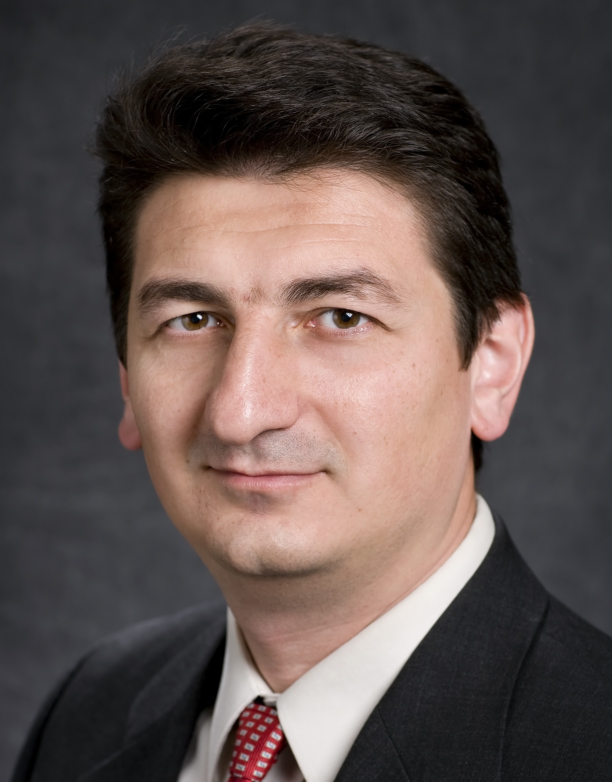}}]{Tevfik Kosar}
is a Professor of Computer Science and Engineering at the State University of New York at Buffalo. He received his Ph.D. in Computer Science from the University of Wisconsin-Madison in 2005. His main research interests include optimization of the performance, scalability, and sustainability of distributed systems, high-performance computing, and big-data analytics pipelines. Some of the awards received by Dr. Kosar include the NSF CAREER Award, IBM Research Award, Google Research Award, IEEE Region-I Technological Innovation Award, UB Senior Faculty Research and Teaching Award, and UB Exceptional Scholar: Sustained Achievement Award. Dr. Kosar is a Senior Member of the IEEE.
\end{IEEEbiography}

\end{document}